\documentclass[aps,pra,superscriptaddress,amsmath,amssymb,reprint]{revtex4-2}

\usepackage{tikz}
\usepackage{graphicx}
\usepackage{dcolumn}
\usepackage{bm}
\usepackage{float}
\usepackage{svg}
\usepackage{braket}
\usepackage{bbold}
\usepackage{enumerate}
\usepackage{wrapfig}
\usepackage{amsmath} 
\usepackage{amssymb} 
\usepackage{seqsplit}
\usepackage{xcolor}
\usepackage{braket}
\usepackage{tabularx}
\usepackage{makecell}
\usepackage[export]{adjustbox}
\usepackage[shortlabels]{enumitem}
\usepackage{setspace}
\usepackage{tensor}
\usepackage{hyperref}

\newcommand{\bes} {\begin{subequations}}
\newcommand{\ees} {\end{subequations}}

\def\theYear{\the\year}
\hypersetup{pdfpagemode=FullScreen, colorlinks=true, allcolors=blue}

\begin{document}

\title{On-demand driven dissipation for cavity reset and cooling}
\author{Vivek Maurya}
\thanks{These two authors contributed equally to this work.}
\affiliation{Center for Quantum Information Science and Technology, University of Southern California, Los Angeles, California 90089}
\affiliation{Department of Physics \& Astronomy, University of Southern California, Los Angeles, California 90089}
\author{Haimeng Zhang}
\thanks{These two authors contributed equally to this work.}
\affiliation{Center for Quantum Information Science and Technology, University of Southern California, Los Angeles, California 90089}
\affiliation{Ming Hsieh Department of Electrical \& Computer Engineering, University of Southern California, Los Angeles, California 90089}
\author{Daria Kowsari}
\affiliation{Center for Quantum Information Science and Technology, University of Southern California, Los Angeles, California 90089}
\affiliation{Department of Physics \& Astronomy, University of Southern California, Los Angeles, California 90089}
\affiliation{Department of Physics, Washington University, St. Louis, Missouri 63130}
\author{Andre Kuo}
\affiliation{Center for Quantum Information Science and Technology, University of Southern California, Los Angeles, California 90089}
\affiliation{Department of Physics \& Astronomy, University of Southern California, Los Angeles, California 90089}
\author{Darian M. Hartsell}
\thanks{Current affiliation: Georgia Tech Research Institute, Atlanta, Georgia 30332}
\affiliation{Center for Quantum Information Science and Technology, University of Southern California, Los Angeles, California 90089}
\affiliation{Department of Physics \& Astronomy, University of Southern California, Los Angeles, California 90089}
\author{Clark Miyamoto}
\thanks{Current affiliation: Department of Physics, New York University, New York, New York 10003}
\affiliation{Center for Quantum Information Science and Technology, University of Southern California, Los Angeles, California 90089}
\affiliation{Department of Physics \& Astronomy, University of Southern California, Los Angeles, California 90089}
\author{Jocelyn Liu}
\thanks{Current affiliation: Department of Electrical and Computer Engineering, Princeton University, Princeton, New Jersey, 08544}
\affiliation{Center for Quantum Information Science and Technology, University of Southern California, Los Angeles, California 90089}
\affiliation{Ming Hsieh Department of Electrical \& Computer Engineering, University of Southern California, Los Angeles, California 90089}
\author{Sadman Shanto}
\affiliation{Center for Quantum Information Science and Technology, University of Southern California, Los Angeles, California 90089}
\affiliation{Department of Physics \& Astronomy, University of Southern California, Los Angeles, California 90089}
\author{Evangelos Vlachos}
\affiliation{Center for Quantum Information Science and Technology, University of Southern California, Los Angeles, California 90089}
\affiliation{Department of Physics \& Astronomy, University of Southern California, Los Angeles, California 90089}
\author{Azarin Zarassi}
\thanks{Current affiliation: Resonant Inc., Goleta, California 93117}
\affiliation{Center for Quantum Information Science and Technology, University of Southern California, Los Angeles, California 90089}
\affiliation{Department of Physics \& Astronomy, University of Southern California, Los Angeles, California 90089}
\author{Kater W. Murch}
\affiliation{Department of Physics, Washington University, St. Louis, Missouri 63130}
\author{Eli M. Levenson-Falk}
\email{elevenso@usc.edu}
\affiliation{Center for Quantum Information Science and Technology, University of Southern California, Los Angeles, California 90089}
\affiliation{Department of Physics \& Astronomy, University of Southern California, Los Angeles, California 90089}
\affiliation{Ming Hsieh Department of Electrical \& Computer Engineering, University of Southern California, Los Angeles, California 90089}

\begin{abstract}
    We present a superconducting circuit device that provides active, on-demand, tunable dissipation on a target mode of the electromagnetic field. Our device is based on a tunable ``dissipator'' that can be made lossy when tuned into resonance with a broadband filter mode. When driven parametrically, this dissipator induces loss on any mode coupled to it with energy detuning equal to the drive frequency. We demonstrate the use of this device to reset a superconducting qubit's readout cavity after a measurement, removing photons with a characteristic rate above $50\ \mu\mathrm{s}^{-1}$. We also demonstrate that the dissipation can be driven constantly to simultaneously damp and cool the cavity, effectively eliminating thermal photon fluctuations as a relevant decoherence channel. Our results demonstrate the utility of our device as a modular tool for environmental engineering and entropy removal in circuit QED. 
\end{abstract}

\maketitle
\section{Introduction}
Superconducting qubits are a promising quantum computing technology, combining fast operation speed, long-lived coherence, and scalability \cite{kjaergaardSuperconductingQubitsCurrent2020}. Great strides have been made in improving qubit coherence through circuit design \cite{paikObservationHighCoherence2011}, materials engineering \cite{oliverMaterialsSuperconductingQuantum2013}, filtering and shielding \cite{barendsMinimizingQuasiparticleGeneration2011}, and novel qubit architecture \cite{nguyenHighCoherenceFluxoniumQubit2019,gyenisExperimentalRealizationProtected2021}. However, relaxation and dephasing still limit processor performance. In addition, readout remains a challenging problem that can even exacerbate the decoherence challenge: environmental coupling required for readout can also be a source of dephasing. Unwanted excitations in the environmental modes can cause ``accidental measurement'' of a qubit state, reducing or destroying phase coherence.


Almost all modern superconducting processors use the circuit quantum electrodynamics (cQED) architecture: a qubit circuit is coupled to a linear resonant circuit or cavity, which in turn is coupled to a microwave feedline \cite{blaisCircuitQuantumElectrodynamics2021}. These circuits are typically operated in the dispersive regime where the detuning $\Delta$ between qubit and cavity modes is far greater than there coupling strength $g$, such that exciting the qubit shifts the resonant frequency of the cavity by the dispersive shift $\chi$. The inverse is also true: the cavity photon number $n$ shifts the qubit frequency by $n \chi$. This creates a possible source of dephasing, as a fluctuating photon number can cause a noisy shift in the qubit frequency~\cite{gambettaQubitphotonInteractionsCavity2006,clerkUsingQubitMeasure2007a,photon-shot-dephasing-schoelkopf}. These fluctuations may be due to residual photons from a coherent drive on the cavity used to perform readout, or thermal photons resulting from the finite temperature of the mode \cite{yanDistinguishingCoherentThermal2018}. In the limit of low average photon number $\bar{n}$ the induced qubit dephasing rate is
\begin{equation} \label{eq:photonDephasing}
    \Gamma_{\phi} \approx \frac{m \chi^2 \kappa}{\chi^2+\kappa^2} \bar{n} 
\end{equation}
where $\kappa$ is the cavity photon loss rate set by its coupling to the environment and $m=1$ for thermal states or $m=2$ for coherent states \cite{clerkUsingQubitMeasure2007a}.
In order to eliminate this dephasing channel, the cavity photon number must be kept very close to zero when a readout is not being performed. This is a challenge, as nearby microwave attenuators heat up and radiate thermal photons into the cavity \cite{yehMicrowaveAttenuatorsUse2017}. Optimized attenuators have successfully damped these photons to the point where their effects are negligible \cite{wangCavityAttenuatorsSuperconducting2019}, but at the cost of severely reduced readout signal to noise ratio due to added attenuation of the signal. Even if the cavity equilibrium temperature were close to zero, after a readout it still takes many times the cavity lifetime for all the photons to leave. Active reset schemes can reduce this ringdown time \cite{mcclureRapidDrivenReset2016,boutinResonatorResetCircuit2017,lledoCloakingQubitCavity2023}, but it remains a rate-limiting step in qubit operation after readout. This issue is especially significant as the field begins to implement fault-tolerant error correction schemes, which require many readout operations per logical operation. Solving these issues requires deterministically resetting \emph{and holding} a cavity in its ground state without diminishing readout fidelity or qubit coherence.

The problem of ensuring that a cavity is in its ground state is essentially one of entropy removal. Entropy can be removed from a system either through active control (i.e., an information-based Maxwell's daemon approach) \cite{brillouinScienceInformationTheory1956,groenewoldProblemInformationGain1971,camatiExperimentalRectificationEntropy2016,pekolaMaxwellDemonBased2016,naghilooInformationGainLoss2018,masuyamaInformationtoworkConversionMaxwell2018,kumarSortingUltracoldAtoms2018,songQuantumProcessInference2021} or through allowing heat to flow to a colder system (i.e., a dissipation-based approach). Information-based approaches are not ideal for removing entropy from a typical readout cavity, as feedback loop delay times (approximately $0.5-5\ \mu\mathrm{s}$) are typically not small compared to $1/\kappa$ (approximately $0.1-1\ \mu\mathrm{s}$). On the other hand, dissipation is in general a valuable resource for control of cQED systems \cite{kapitUpsideNoiseEngineered2017a,harringtonEngineeredDissipationQuantum2022}. In the cQED architecture, dissipation can be tuned fast on-demand by using driven couplings between a target mode and a lossy bath. This approach has been used for qubit reset \cite{valenzuelaMicrowaveInducedCoolingSuperconducting2006,geerlingsDemonstratingDrivenReset2013,magnardFastUnconditionalAllMicrowave2018,eggerPulsedResetProtocol2018,zhouRapidUnconditionalParametric2021a}, error reduction \cite{lacroixFastFluxActivatedLeakage2023}, preparation and stabilization of single qubits \cite{murchCavityAssistedQuantumBath2012,luUniversalStabilizationParametrically2017a}, entangled states \cite{shankarAutonomouslyStabilizedEntanglement2013,kimchi-schwartzStabilizingEntanglementSymmetrySelective2016a}, and stabilization of bosonic encodings \cite{hollandSinglePhotonResolvedCrossKerrInteraction2015, leghtasConfiningStateLight2015}. Here we apply on-demand dissipation to remove entropy from a readout cavity.

In this paper we present a modular dissipative element that can be used to induce tunable, on-demand loss in a target mode. Our ``dissipator'' consists of a tunable coupler, similar to a high-frequency tunable transmon, that links a target system to a lossy transmission line via a bandpass filter, similar to the geometry proposed in \cite{wongTunableQuantumDissipator2019}. By driving the dissipator parametrically, we can cause the target mode to hybridize with the dissipator and thus become lossy itself. We demonstrate the utility of this device by using it to remove photons from the readout cavity of a standard superconducting cQED system. We show that using the dissipator it is possible to damp the cavity with a characteristic rate greater than $50\ \mu\mathrm{s}^{-1}$ and achieve a complete reset after measurement in 170 ns, more than 10 times faster than the background decay rate. We also show that the dissipator can be used to cool the cavity during qubit operation, reducing the photon-induced decoherence rate on the qubit to the point that it is negligible. Our results show the utility of the dissipator for reset and cooling operations, and as a general source of on-demand loss.

\section{Principle of Operation}
The operation of our device is based on a parametrically-driven exchange interaction (also sometimes referred to as a swapping or beamsplitter interaction \cite{luHighfidelityParametricBeamsplitting2023}) between a target mode and a lossy mode. If the lossy mode is near its ground state and the loss is fast compared to the exchange (i.e., the exchange is overdamped), this is effectively a one-way interaction: excitations in the target mode are shuttled to the lossy mode, where they dissipate before they can swap back into the target. Crucially, the lossy mode and target mode can be far detuned, as the parametric drive allows for frequency conversion. This prevents static coupling from causing unwanted loss on the target mode when no parametric drive is present. We can therefore tune the strength of the loss on the target simply by tuning the strength of the parametric drive, turning it on and off as needed with broad bandwidth.

\begin{figure}
    \centering
    \includegraphics{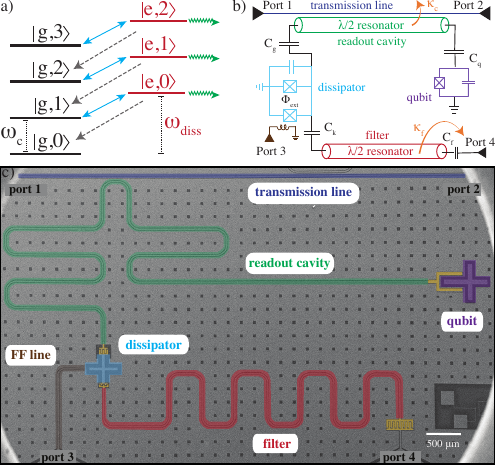}
    \caption{\label{fig:circuit_diagram}
    (a) Energy level diagram depicting the driven dissipation process, showing the lowest two dissipator states ($\ket{g},\ket{e}$) and lowest four cavity photon states ($\ket{0},\ket{1},\ket{2},\ket{3}$). A parametric drive on the dissipator at the dissipator-cavity detuning frequency induces coherent oscillations between the $\ket{g,n} \leftrightarrow\ket{e,n-1}$ states (blue arrows). The dissipator is lossy, and so quickly emits a photon (green) and decays back to the $\ket{g,n-1}$ state (black dashed arrow). A photon is thus unconditionally removed from the cavity; continuously driving this process brings the system to its ground state $\ket{g,0}$. (b) Circuit schematic of our device. A measurement feedline (black, ports 1 and 2) is coupled with rate $\kappa_c$ to a half-wave cavity (green) used to read out an ordinary transmon qubit (purple). The cavity is also coupled to a tunable coupler, the \emph{dissipator} (light blue), which can be flux driven via a fast flux (FF) line (port 3). The dissipator is coupled via a bandpass filter (implemented as a half-wave cavity) to a terminated transmission line (port 4), with filter bandwidth $\kappa_f$, causing the dissipator to become lossy when near the filter band. (c) False-colored scanning electron microscope image of a device with the same design as the one reported here, with color coding as used in (b). 
    The readout cavity, qubit, dissipator, filter, and terminated line are all connected via coupling capacitors (highlighted in yellow).
    }
\end{figure}

\subsection{Parametrically-Driven Dissipation}\label{sec:parametricTheory}

We begin by considering the case of a lossy dissipator which is coupled to a lossless linear oscillator. For simplicity, we treat the dissipator as a tunable qubit with frequency $\omega_\mathrm{diss}$.
The dissipator is coupled to a cavity of frequency $\omega_\mathrm{c}$ with coupling strength $g_c$. This gives the familiar Jaynes-Cummings Hamiltonian\begin{align}\label{eq:hamiltonian}
	H=\omega_\mathrm{c} a^{\dagger} a-\frac{\omega_\mathrm{diss}}{2} \sigma_z + g_\mathrm{c} \left(a^{\dagger}+a\right)\left(\sigma_{+}+\sigma_{-}\right),
\end{align}
where $\sigma_+$ ($\sigma_-$) is the raising (lowering) operator for the dissipator, and $a$ ($a^\dagger$) is the annihilation (creation) operator for the cavity. 

We next turn on a parametric drive which periodically modulates the dissipator frequency \begin{align}\label{eq:drive}
	H_{p}(t)= \epsilon_\mathrm{p}  \sin \left(\omega_\mathrm{p} t\right) \sigma_z,
\end{align}
where $\epsilon_\mathrm{p}$ is the parametric drive amplitude and $\omega_\mathrm{p}$ is the parametric drive frequency. 
This parametric modulation of the dissipator transition frequency causes sidebands to appear, centered on $\omega_\mathrm{diss}$ and spaced by $\omega_\mathrm{p}$. Parametric coupling can be turned on by tuning one of these sidebands into resonance with the cavity. Typically the first sideband is used, i.e., $\omega_\mathrm{p}= \omega_\mathrm{c}-\omega_\mathrm{diss}\equiv \Delta$. This resonance results in a parametrically-driven exchange interaction between the cavity mode and the lossy dissipator mode.
In the dispersive regime where $g_c\ll \Delta$, and when the drive amplitude is small $\epsilon_\mathrm{p}\ll\Delta$ \footnote{This is the regime in which we operate our device, based on the measured parameters. Note that if we were not in this regime, during modulation the dissipator frequency would come near resonance with the cavity, and the two would undergo a direct exchange operation. This may produce the desired loss, but such frequency collisions are exactly what we sought to avoid with parametric drives, as there may be other modes near the cavity that we do not wish to make lossy.}, the drive-induced parametric coupling rate is given by 
\begin{align}
\label{eq:gp}
    g_\mathrm{p} \approx \frac{g_\mathrm{c} \epsilon_\mathrm{p}}{\Delta}
\end{align}
leading to Rabi oscillations between the states $\ket{g,n+1}\leftrightarrow\ket{e,n}$ with frequency
\begin{align}\label{eq:swap_rate}
	\Omega_R =  \frac{\sqrt{n+1} g_\mathrm{c} \epsilon_\mathrm{p}}{\Delta}.
\end{align}
We derive and generalize these equations in a closed-system setting in Appendix \ref{app:parametric}. 

The dissipator loss damps the Rabi swapping oscillations between cavity and dissipator. Depending on whether the parametric coupling $g_\mathrm{p}$ is larger than, smaller than, or equal to the amplitude damping rate $\kappa_\mathrm{diss}/4$, the interaction will be underdamped, overdamped, or critically damped, respectively. In the underdamped case, excitations swap back and forth between dissipator and cavity, and decay at a rate $\kappa_\mathrm{diss}/2$. In the critically-damped case the decay is the same, but no oscillations occur. In the overdamped case the decay is rate-limited by $g_\mathrm{p}$, and the population decays with rate 
\begin{equation}\label{eq:lossrate}
\kappa_\mathrm{eff} = \frac{1}{2} \left(\kappa_\mathrm{diss}-\sqrt{\kappa_\mathrm{diss}^2-16 g_\mathrm{p}^2}\right).
\end{equation}
In the small-coupling limit ($g_\mathrm{p}\ll \kappa_\mathrm{diss}$) the decay rate is roughly $\kappa_\mathrm{eff} \approx 4g_\mathrm{p}^2 / \kappa_\mathrm{diss}$. These rates are obtained in \cite{zhouRapidUnconditionalParametric2021a} using non-Hermitian Hamiltonians and solving for the explicit time dynamics; here we quote the result rather than reproducing the lengthy analysis. We note that, once the parametric coupling rate is known, the damping rate could also easily be obtained treating the cavity and dissipator as classical damped oscillators. It is also possible to derive the parametric coupling rate classically, but the quantum derivation is far simpler mathematically \cite{okamotoCoherentPhononManipulation2013}.


\subsection{Refrigeration}\label{sec:refTheory}
In this section we describe how the dissipator can be operated as a quantum refrigerator and give a brief derivation of the equilibrium behavior. The refrigeration performance of similar systems has been explored theoretically in \cite{wongTunableQuantumDissipator2019,doucetHighFidelityDissipation2020} and experimentally in \cite{zhouRapidUnconditionalParametric2021a}, among others. Here we give a simplified summary in the relevant parameter regimes. A fuller description without simplifying assumptions is in Appendix \ref{app:refrigeration}.

When the dissipator is tuned near resonance with the filter, the exchange interaction between them allows heat to flow. Since loss via the filter dominates all other sources of loss on the dissipator, in the absence of other drives the dissipator comes to thermal equilibrium with the filter. The filter loss is likewise dominated by the coupling to the thermal bath formed by the continuous spectrum of the lossy feedline, and so the filter is in thermal equilibrium with this bath. Therefore $T_\mathrm{diss}\approx T_\mathrm{bath}$.

On the other hand, the target (cavity) mode is far off resonance from the dissipator and the two cannot easily exchange heat. Therefore the cavity and dissipator do \emph{not} reach thermal equilibrium. Instead they can exchange excitations via the coherent swapping interaction induced by the parametric drive. In the absence of any drive, the cavity comes into equilibrium with its environment at temperature $T_0$ with a rate $\kappa_\mathrm{c}$. When the drive is turned on, excitations are exchanged between the cavity and dissipator. In the underdamped limit where these swaps happen very fast compared to the cavity and dissipator loss rates ($g_p\gg \kappa_\mathrm{c},\kappa_\mathrm{diss}$), the cavity and dissipator reach an equilibrium of \emph{excitation number}: $\bar{n}_\mathrm{c} = \bar{n}_\mathrm{diss}$. In the limit where the dissipator's loss dominates $\kappa_\mathrm{diss}\gg\kappa_\mathrm{c}$, the dissipator comes to the temperature of its environment $T_\mathrm{bath}$. The mean photon number is given by 
\begin{equation}\label{eq:photonNumTherm}
\bar{n} = \frac{1}{e^{\hbar \omega_\mathrm{diss} / k_B T_\mathrm{bath}}-1} = \frac{1}{e^{\hbar \omega_\mathrm{c} / k_B T_\mathrm{c}}-1}
\end{equation}
where $T_\mathrm{c}$ is the cavity's effective temperature under driving and $k_B$ is the Boltzmann constant. This gives $T_\mathrm{c} = \frac{\omega_\mathrm{c}} {\omega_\mathrm{diss}} T_\mathrm{bath}$. Thus if $\frac{\omega_\mathrm{c}} {\omega_\mathrm{diss}} T_\mathrm{bath} < T_0$ the cavity is cooled by the dissipation drive.

The opposite limit is also interesting: the overdamped regime where the cavity-dissipator swapping is slow compared to the dissipator loss rate ($g_p \ll \kappa_\mathrm{diss}$). This is the regime in which we typically operate our device. Here, we can treat the target cavity as if it is exposed to two baths: the normal background bath with temperature $T_0$ and coupling rate $\kappa_\mathrm{c}$, set mainly by the cavity coupling to the measurement line; and the effective thermal bath formed by the frequency-converted excitations from the dissipator with driven coupling (loss) rate $\kappa_\mathrm{eff} \approx g_p^2 / \kappa_\mathrm{diss}$ (from Sec. \ref{sec:parametricTheory}). This latter bath has effective temperature $T_\mathrm{eff} = \frac{\omega_\mathrm{c}}{\omega_\mathrm{diss}} T_\mathrm{bath}$, which can be derived as above by counting excitations in the dissipator and inverting Eq.~\ref{eq:photonNumTherm} to find $T_\mathrm{eff}$. This leads to an overall cavity temperature 
\begin{equation}\label{eq:cavTemp}
T_\mathrm{c} = \frac{\kappa_\mathrm{c} T_0 + \kappa_\mathrm{eff} \frac{\omega_\mathrm{c}} {\omega_\mathrm{diss}} T_\mathrm{bath}}{\kappa_\mathrm{c}+\kappa_\mathrm{eff}}. 
\end{equation}
Once again we can lower the cavity temperature so long as $\frac{\omega_\mathrm{c}} {\omega_\mathrm{diss}} T_\mathrm{bath} < T_0$, which is easily achieved by setting $\omega_\mathrm{diss} > \omega_\mathrm{c}$ if the baths are at similar temperatures. 


\section{Experimental Results}
\subsection{Device and Apparatus}
A circuit schematic and image of our device are given in Figure \ref{fig:circuit_diagram}. Our device consists of a few modular elements. First is an ordinary cQED readout system, with a fixed-frequency transmon qubit at $\omega_q/2\pi = 3.368$ GHz, with anharmonicity of $\alpha_\mathrm{q}/2\pi = -172$ MHz, coupled to one end of an open-ended section of co-planar waveguide (CPW) that forms a half-wave cavity at $\omega_\mathrm{c}/2\pi = 5.594$ GHz. The qubit and cavity couple with a rate $g_q/2\pi = 53.9$ MHz, and the cavity couples to the external measurement line at rate $\kappa_\mathrm{c}/2\pi = 500$ kHz. The second element is a tunable coupler with flux-tunable resonant frequency $\omega_\mathrm{diss}(\phi) \in [4.2,15.3]$ GHz, which couples to the opposite end of the cavity with rate $g_\mathrm{c}/2\pi \approx 145$ MHz; we term this the ``dissipator''. A fast flux line allows us to drive microwave flux tones into the dissipator SQUID loop, modulating its energy. The third element is our source of loss, which is implemented as a transmission line terminated with a 50 $\Omega$ load. The load is anchored to the base plate of the cryostat at 7 mK. Between the dissipator and load is another half-wave cavity at frequency $\omega_\mathrm{f}/2\pi = 8.6$ GHz, which couples to the dissipator with rate $g_\mathrm{f}/2\pi = 535$ MHz and to the external feedline with rate $\kappa_\mathrm{f}/2\pi \approx 120$ MHz. We term this the ``filter'' mode as it forms a bandpass filter that suppresses static coupling between the cavity and the terminated feedline, avoiding unwanted loss when the dissipator is not driven. The filter linewidth is approximate as it depends on the static flux bias (i.e., on the dissipator frequency) and has a non-Lorentzian lineshape due to impedance variations over the broad bandwidth. The dissipator-cavity coupling is also approximate due to the fact that the effective coupling rate is modified by the filter when the two are near resonance. Device parameters are extracted as described below; more details can be found in Appendix \ref{app:calib}.

\begin{figure}
    \centering
    \includegraphics{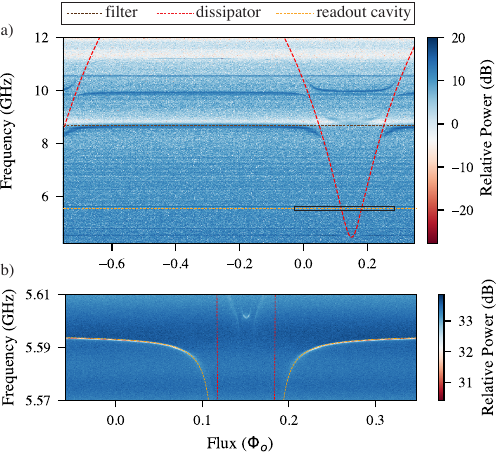}
    \caption{\label{fig:fluxtuning}
    (a) Spectroscopy of cavity and filter modes as a function of flux through the dissipator SQUID loop. The upper plot shows power transmitted from the filter (loss) port to the measurement output port (port 4 to 2), relative to an arbitrary reference power in units of dB. The dashed red line is a fit to the dissipator frequency as a function of flux based on the locations of the avoided crossings. The broadband mode at $8.6$ GHz is the filter mode (dashed brown line). The fundamental cavity mode is indicated by the dashed orange line at $5.594$ GHz. Also visible are the second cavity mode at $11.2$ GHz and another mode at $10$ GHz that we attribute to nonlinear mixing of second cavity and filter modes; the second cavity mode is too narrow to see on this scale. (b) Detailed measurement of the region indicated by the box in (a). Here we show transmission through the ordinary measurement feedline (port 1 to 2), which couples strongly to the readout cavity mode. The red curves which appear vertical are actually the (steep) dissipator flux tuning in this range, and orange curves are the cavity frequency using the dissipator parameters from the previous fit and letting $\omega_\mathrm{c}, g_\mathrm{c}$ be free parameters. }
   
\end{figure}

We tune up our device with spectroscopic measurements. We find the cavity modes by measuring the magnitude of the transmission from the measurement input port (port 1) to the output port (port 2) and looking for features in the transmission spectrum. We perform this measurement with a constant cavity drive. The transmitted signal is amplified, demodulated to 200.25 MHz in a standard heterodyne setup, and digitized. Ordinarily the transmission line that provides the lossy port (port 4) would be terminated; however, to better characterize our device, we instead connect it to a heavily-attenuated drive line. The microwave attenuators on the line produce a nearly identical load to a termination, so we expect the device behavior to be unaffected. See Fig.~\ref{fig:fridge diagram} for an experimental wiring diagram. We can thus characterize the filter mode by measuring transmission from this ``loss port'' (4) to the measurement output port (2), in this case with a vector network analyzer (VNA). The VNA is removed for all other measurements, and the line is terminated at room temperature. The qubit frequency can be found with ordinary pulsed spectroscopy. However, the dissipator is so lossy that it is difficult to see with pulsed spectroscopic measurements. Instead, we measure the cavity and filter spectra while tuning the dissipator flux, and fit the dissipator frequency $\omega_\mathrm{diss}(\phi)$ and coupling rates $g_\mathrm{c},g_\mathrm{f}$ based on the observed avoided crossings. See Figure \ref{fig:fluxtuning}. We observe crossings with the fundamental and second cavity modes at $5.59$ and $11.2$ GHz, the filter mode at $8.6$ GHz, and a spurious feature at $9.9$ GHz that we attribute to a nonlinear mixing process where 2 measurement photons convert into 1 filter photon and 1 photon in the 2nd mode of the cavity.

We note that it is possible to operate with the dissipator tuned to any frequency by parametric driving at the detuning between the cavity and filter modes, inducing an exchange interaction directly between the cavity and filter without populating the dissipator. Indeed, we measured increased cavity loss when driving the dissipator at the cavity-filter detuning at all static flux biases. However, we found the strongest parametrically-driven loss when the dissipator was resonant with the filter, and so we restrict our results and analysis to that regime. The filter resonance lineshape changes enough that we are unable to fit the linewidth (i.e., decay rate) when the dissipator is on resonance with the filter, and so the precise dissipator decay rate $\kappa_\mathrm{diss}$ is unknown. We expect $\kappa_\mathrm{diss} \approx \kappa_\mathrm{f}/2$, but we can only bound $\kappa_\mathrm{diss}/2\pi \in [10,100]$ MHz with measurements.

\subsection{Cavity Ringdown}

\begin{figure}
    \centering
    \includegraphics{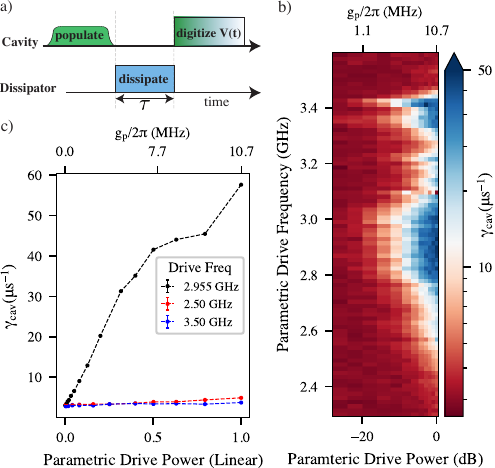}
    \caption{\label{fig:ringdown}
    (a) Pulse sequence for cavity ringdown measurement. The cavity is populated with an initial drive pulse. We then drive the dissipator with a parametric flux drive for a variable time $\tau$ before digitizing the signal resulting from the remaining cavity population. (b) Ringdown spectroscopy measurement of cavity ringdown rate as a function of parametric drive frequency and power. The broad feature centered at 2.9 GHz is the detuning between cavity and dissipator. Narrower features at $\sim 3.35$ and $\sim 3.2$ GHz are spurious signals due to crosstalk-induced driving of the qubit transitions. (c) Horizontal slices of the data in (b) at 3 different frequencies. Only when the parametric drive is on resonance with the dissipator-cavity detuning (black curve) do we see a strong enhancement of the ringdown rate. The upper horizontal axis, which reports the drive in units of induced parametric coupling rate, is based on estimates of $g_c$ and $\epsilon_p$ and should be viewed as an estimate itself; it roughly predicts the decay rate seen.
    }
\end{figure}

We first characterize the cavity-dissipator interaction by measuring cavity ringdown with the parametric drive off. See Figure \ref{fig:ringdown}. We displace the cavity with a coherent drive pulse, populating with a mean photon number $\bar{n}\approx 39$ similar to a normal readout pulse. We then wait a variable time $\tau$ and then digitize the signal for a short time of $t_m = 32$ ns. The signal is a two-channel heterodyne voltage, as above. While we collect the full vector cavity response, we focus on the cavity output power. We take many averages, thus finding the average cavity output power in the short interval $t = [\tau,\tau+t_m]$. We sweep $\tau$ and fit the resulting exponential decay of the output power to extract the cavity ringdown rate $\gamma_\mathrm{cav}$ 
which should equal $\kappa_\mathrm{c}$ when no dissipation is driven. 
At the operating point we chose, the cavity rings down with a rate $\gamma_\mathrm{cav} = 3.000 \pm 0.036\ \mu\mathrm{s}^{-1} \approx \kappa_\mathrm{c}$ when the dissipator is not driven. Note that we are characterizing ringdown of the cavity output power, which is linearly proportional to the photon number, not the output amplitude as is often done. The ringdown rate for amplitude decay is half the rate for power decay. The qubit is left idle during this measurement, and remains in its ground state.

We next add a parametric drive at the cavity-dissipator detuning frequency during the ringdown, inducing driven dissipation. Again we leave the qubit in its ground state for the entire measurement. We measure the ringdown time as a function of the amplitude and frequency of this parametric drive. This ``ringdown spectroscopy'' shows a strongly enhanced ringdown rate when the parametric drive frequency $\omega_\mathrm{p}/2\pi \approx (\omega_\mathrm{diss}-\omega_\mathrm{c})/2\pi = 2.95$ GHz, as expected. When driven with a moderate parametric drive amplitude of $\approx 0.005$ $\Phi_o$ (corresponding to a frequency modulation $\epsilon_p/2\pi \approx 130$ MHz), we are able to increase the ringdown rate to $\gamma_\mathrm{cav} = 57.4 \pm 1.5$ $(\mu\mathrm{s})^{-1}$. The parametric drive amplitude is approximate as we do not know the precise attenuation of the flux line at this frequency. If we assume $\kappa_\mathrm{diss} \approx \kappa_\mathrm{f}/2$ when the dissipator is near resonance with the filter, we have a predicted decay rate from Eq.~\ref{eq:lossrate}: $\gamma_\mathrm{cav} = \kappa_{eff} \approx 60\ \mu\mathrm{s}^{-1}$, in good agreement with the measurement. The approximately linear relationship between $\gamma_\mathrm{cav}$ and parametric drive power (i.e., $g_\mathrm{p}^2$) is also in agreement with Eq.~\ref{eq:lossrate}.
We note that the dissipator-induced loss rate simply adds to the bare cavity loss rate, and so even a very high-Q cavity could see a similar ringdown rate when the dissipator is driven.

We see additional spurious features corresponding to transitions of the fixed-frequency transmon or combined transmon-cavity system. For instance, the features near $\omega_\mathrm{p}/2\pi \approx 3.3$ GHz are the transmon transitions $\omega_{ge}$ and $\omega_{gf}/2$, broadened due to the strong drive. These features are due to classical crosstalk from the drive line to the transmon, as our parametric flux drive also induces a small charge drive on the transmon. This changes the transmon state and thus the cavity frequency, leading to a ringdown at a different frequency than the initial cavity drive. This signal is filtered by our measurement electronics and so appears as a faster ringdown. Only the true cavity-dissipator exchange feature leads to removal of photons from the cavity, as we show below. There is an additional sharper feature at $3.4$ GHz which we do not know the origin of, but appears to cause no real photon removal and so is ignored in our analysis.

The measured ringdown rate increases with parametric drive power up to the highest powers we are able to drive, as expected in the overdamped regime. At high enough drive power we would expect the ringdown rate to saturate at $\gamma_\mathrm{cav} = \kappa_{diss}/2$. At drive powers roughly 20 dB above our maximum drive, the dissipator frequency would cross the cavity frequency during modulation, and ordinary resonant exchange would occur. This could lead to significant damping, but we choose not to analyze this regime as it would also lead to frequency collisions between the dissipator and any mode between its mean frequency and the cavity frequency. Avoiding such collisions is the main motivation for using parametric driving. 

We note that the dissipator has significant anharmonicity---we estimate $\alpha_\mathrm{diss}/2\pi \approx -350$ MHz from finite-element simulations of the dissipator capacitance. As such, when the dissipator is populated with a photon, the parametric drive is detuned from the frequency necessary to swap a cavity photon into the dissipator's higher levels. We would expect to see this photon blockade only if the anharmonicity was larger than both the photon swap rate (which grows with photon number) and the dissipator decay rate. Additionally, the blockade would only appear if the photon swap rate was itself larger than the dissipator decay rate; otherwise, the mean photon number in the dissipator would remain low. If photons were blockaded from leaving the cavity, we would expect to see a non-exponential ringdown. Our ringdown measurements are well fit by exponential curves (to within the accuracy of our measurements), and so we see no effect of dissipator anharmonicity in our device.

\subsection{Reset After Measurement}
\begin{figure}
    \centering
    \includegraphics{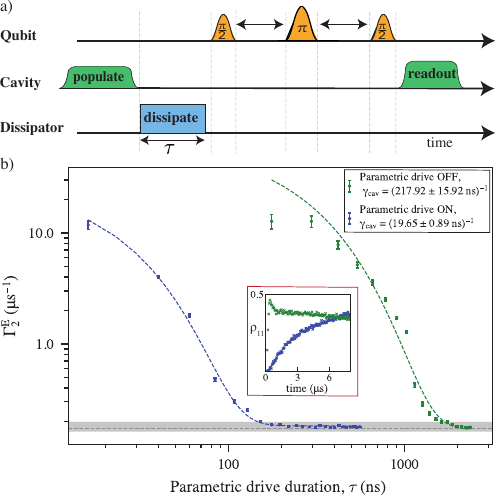}
    \caption{\label{fig:cavityReset}
    (a) Pulse sequence for cavity reset measurments. We first drive a population pulse to displace the cavity state. We then drive a parametric pulse on the dissipator for a variable time to induce cavity loss. We next perform an ordinary Hahn echo sequence on the qubit and then measure its state. (b) Measured echo decoherence rate $\Gamma_2^E$ as a function of the wait time after an initial population pulse, with (blue, left) and without (green, right) the dissipator drive. Decoherence initially is fast, then drops to the background value (gray shaded region, indicating the variance in background $\Gamma_2^E$). The dashed curves are fits to the simple model given in Eq.~\ref{eq:resetDephasing}, with extracted cavity decay rates given in the legend. 
   At short times without dissipator drive, the photon number changes significantly during the echo measurement, making $\Gamma_2^E$ fits unreliable. This can be seen clearly in the inset curves, which show single echo measurements with (blue) and without (green) the dissipation drive: the measurement without dissipation drive gives a strongly non-exponential curve. 
    }
\end{figure}

We next demonstrate how the dissipator can be used to reset the readout cavity after a measurement. The pulse sequence is shown in Figure $\ref{fig:cavityReset}$(a). We populate the cavity with an ordinary readout pulse as in the ringdown measurements above, putting $\bar{n} \approx 39 $ photons in the cavity. We next wait 80 ns to ensure no pulses inadvertently overlap due to mismatches in signal propagation length. We then drive the dissipator parametrically for a variable time $\tau$. After tuning off this drive, we perform a standard Hahn echo sequence measurement on the qubit to characterize the dephasing induced by cavity photons. 
We also measure coherence with the cavity population pulse but without the parametric drive to characterize the bare cavity reset time.

The background $\Gamma_2^E$ we measure varies in a small range about $0.17\ \mu \mathrm{s}^{-1}$, with the typical 1-standard-deviation range shown by the shaded bar in Fig.~\ref{fig:cavityReset}. When the cavity is populated and there is a short delay (between $0.2-1$ $\mu$s) before starting the echo sequence with no parametric reset drive, $\Gamma_2^E$ is very fast (between $2-10\ \mu\mathrm{s}^{-1}$). We put low confidence on the exact fast dephasing rate at the shortest wait times, as the decay curves become non-exponential (see Fig.~\ref{fig:cavityReset}(b) inset), as is expected since the cavity photon number changes during the qubit evolution. Indeed, the echo pulse sequence itself may be miscalibrated with so many photons in the cavity. 
After roughly $2.2$ $\mu$s, or just under 7 times the bare cavity ringdown time of $333$ ns, the coherence recovers to its original value. In contrast, when we drive a reset pulse on the dissipator flux line with a frequency of $2.955$ GHz and amplitude $g_\mathrm{p}\approx 11$ MHz (giving $\kappa_\mathrm{eff} = \gamma_\mathrm{cav} = 57\ \mu\mathrm{s}^{-1})$, we are able to recover the qubit coherence in just $\approx 170$ ns, roughly 10 times the driven ringdown time $1/\gamma_\mathrm{cav} = 17.36 \pm 0.48$ ns. We are thus able to quickly reset the readout cavity and rapidly eliminate the cavity-photon-induced dephasing after a readout. The reset speedup is roughly equal to the cavity ringdown rate speedup, as expected.

We can fit the qubit decoherence rate during cavity reset to a simple functional form:
\begin{equation}\label{eq:resetDephasing}
\Gamma_{2}(\tau) = \frac{2\bar{n} \chi^2 \kappa_\mathrm{c}}{\chi^2+\kappa_\mathrm{c}^2} e^{-\gamma_\mathrm{cav} \tau} + \Gamma_{2}^0
\end{equation}
where the first term describes dephasing induced by a decaying coherent state and $\Gamma_{2}^0$ is the bare qubit decoherence rate in the absence of additional cavity population. 
We note that this is a simplified model that does not include the full open system dynamics of the qubit-cavity-bath interaction. However, it still fits the data quite well in the driven case, and fits the undriven case well after approximately one cavity decay time, as shown in Fig.~\ref{fig:cavityReset}(b). We leave $\gamma_\mathrm{cav}$, $\bar{n}$, and $\Gamma_2^0$ as free parameters. We extract a value of $\gamma_\mathrm{cav} = 50.9 \pm 2.29\ \mu\mathrm{s}^{-1}$
in the driven case, in reasonable agreement with the value $\gamma_\mathrm{cav} = 57.4 \pm 1.5\ \mu\mathrm{s}^{-1}$ 
measured in driven ringdown. We extract $\bar{n} = 39.8\pm 5.5$ and $\Gamma_2^0 = 0.18 \pm 0.002\ \mu\mathrm{s}^{-1}$, also in agreement with prior measured values. In the undriven case the fit value $\gamma_\mathrm{cav} = 4.61 \pm 0.30\ \mu\mathrm{s}^{-1}$
disagrees with the measured $\gamma_\mathrm{cav} = 3.000 \pm 0.036\ \mu\mathrm{s}^{-1}$
from ringdown. We attribute this discrepancy to the fact that the cavity photon population changes significantly through the entire echo sequence, and so the effective $\Gamma_2$ is not a constant; this is confirmed by the highly non-exponential coherence decay shown in the inset of the Fig.~\ref{fig:cavityReset}(b). 
Nevertheless, we can quantitatively say that the driven reset allows the qubit to recover its background coherence in $\sim 170$ ns, while the undriven case takes more than $2000$ ns. This compares favorably with pulse-shaping protocols for resonator reset, which typically speed reset by a factor of $\sim 2$ \cite{mcclureRapidDrivenReset2016}; in our case, the reset speedup is additive, not multiplicative, and so our reset works even for cavities with low intrinsic $\kappa_\mathrm{c}$.

In order to be used as a practical cavity reset, it is essential that the dissipator not compromise readout fidelity. We have performed single-shot readout fidelity with and without a cavity population pulse and reset pulse beforehand, and found no significant change in the average readout fidelity of 0.81. This fidelity is limited by the finite lifetime of our qubit and nonideal performance of our readout amplification chain, which has not been optimized for the cavity frequency.

\subsection{Cavity Refrigeration}
As described above in Sec.~\ref{sec:refTheory}, it is possible to use the dissipator to cool a target system by pumping heat from the system into the dissipator's bath. Importantly, this pumping is frequency-selective, and thus can be used to cool one mode while leaving other modes unaffected. Thus, we can use the dissipator to remove thermal photons from the cavity, cooling it closer to its ground state, while leaving the qubit coherence unaffected.

\begin{figure}
    \centering
    \includegraphics{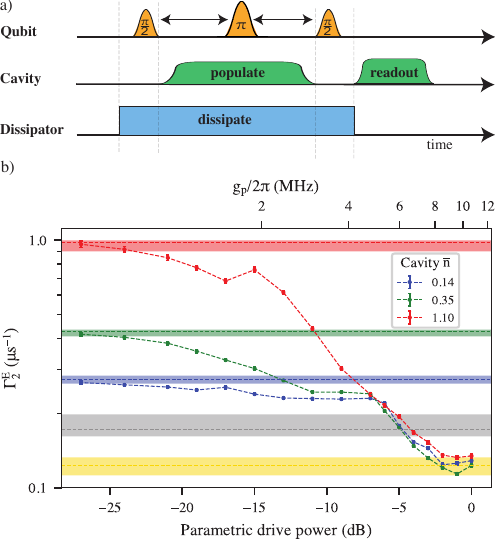}
    \caption{\label{fig:cavity cooling}
    (a) Pulse sequence for the cavity refrigeration measurements. The dissipator is parametrically driven to induce cavity loss while an ordinary Hahn echo sequence is performed on the qubit. During the free evolution of the echo, the cavity is driven with a variable-amplitude population drive to add a fluctuating photon population. (b) Measured echo decoherence rate $\Gamma_2^E$ as a function of parametric drive power for 3 different cavity drive powers. The legend shows the mean cavity photon numbers $\bar{n}$ under the influence of the cavity drives when the parametric (dissipation) drive is off. The gray dashed line is the mean value of $\Gamma_2^E = 0.17\ \mu\mathrm{s}^{-1}$ in the absence of all other drives. The rate fluctuates, and the gray shaded region shows the 1 standard deviation interval. When a cavity drive is present with no cooling drive, fluctuating photon populations of $\bar{n} = 0.14, 0.35$, and $1.10$ in the cavity lead to dephasing, giving the increased average decoherence rates of $0.274\ \mu\mathrm{s}^{-1}$, $0.425\ \mu\mathrm{s}^{-1}$, and $0.980\ \mu\mathrm{s}^{-1}$ shown by the blue, green, and red dashed lines, respectively. Shaded regions indicate 1 standard deviation intervals. When the dissipator is coupled to the cavity with a parametric drive, these excitations are rapidly removed from the cavity and the coherence is improved above its background value. With optimal dissipator driving and no added cavity drive, we read an average decoherence rate of $0.124 \ \mu\mathrm{s}^{-1}$, indicated by the yellow dashed line and shaded region.
    }
\end{figure}

We measure this cavity refrigeration technique with a modified echo sequence. See Figure \ref{fig:cavity cooling}. We first drive an ordinary echo sequence on the qubit as shown in Fig.~\ref{fig:cavity cooling}(a). With no other drives present, we see the ``bare'' $\Gamma_2^E\approx 0.17\ \mu\mathrm{s}^{-1}$ also seen in the cavity reset measurements. We then deliberately degrade this coherence by driving a weak coherent tone on the cavity during the free evolution. This populates the cavity with a fluctuating photon number, dephasing the qubit and raising $\Gamma_2^E$ to approximately $1\ \mu\mathrm{s}^{-1}$ for a population drive that creates a mean photon number $\bar{n}\approx 1$. Finally, we add the parametric drive during the entire sequence, damping and cooling the cavity. For all values of the cavity drive, adding the dissipator parametric drive reduces the $\Gamma_2^E$ observed, enhancing coherence. For sufficiently high parametric drive amplitudes, we can decrease $\Gamma_2^E$ even below its bare value, indicating that thermal photons were likely a source of dephasing even without the cavity drive. When we turn off the cavity drive and drive only the parametric cooling tone, we reach a minimum $\Gamma_2^E = 0.114\ \mu\mathrm{s}^{-1}$. All other qubits made with our in-house fabrication process, which is limited by oxide contaminants in the qubit capacitor, show dephasing rates at or above this level, even when they are well-isolated from their measurement cavities. We therefore suspect the remaining dephasing is due to other processes and not thermal photon noise. This assumption is supported by the fact that the bare dephasing rate and the dephasing rate during refrigeration both drift in time, but the difference between them is relatively constant, indicating it is likely due to other time-varying processes as has been ubiquitously observed in superconducting qubits. Using the mean ``cooled'' value as the dephasing rate due to other processes $\Gamma_2^0 = 0.124$ $\mu$s$^{-1}$ and comparing to the mean bare dephasing rate $\Gamma_2 = 0.172$ $\mu$s$^{-1}$, we can extract a photon-induced dephasing rate $\Gamma_\phi^p = 0.048$ $\mu$s$^{-1}$. To convert this to a mean cavity photon number we can use Eq.~\ref{eq:photonDephasing} along with the measured device parameters $\kappa_c/2\pi = 477$ kHz, $\chi/2\pi = 200$ kHz. We find $\bar{n} \approx 0.107$, which gives an effective cavity temperature $T_\mathrm{c} \approx 115$ mK before refrigeration. 
Assuming $T_\mathrm{diss} \approx T_\mathrm{c}$ and using the bare $\kappa_\mathrm{c} = 3.00$ $\mu$s$^{-1}$ and dissipator-induced $\kappa_\mathrm{eff} = \gamma_\mathrm{cav}-\kappa_\mathrm{c} = 54$ $\mu$s$^{-1}$ (both measured from cavity ringdown), we use Eq.~\ref{eq:cavTemp} to calculate a final cavity temperature $T_\mathrm{c} \approx 77$ mK during refrigeration. This gives an expected photon number $\bar{n} = 0.032$ and an expected dephasing rate $\Gamma_\phi^p \approx 0.9$ ms$^{-1}$. The dephasing is suppressed both due to the reduction of $\bar{n}$ (cooling) and the enhancement of total cavity loss rate $\kappa = \kappa_\mathrm{c}+\kappa_\mathrm{eff}$ such that $\kappa \gg \chi$, reducing the dephasing rate per photon. Therefore, with active refrigeration of the cavity mode, the limit on $T_2$ due to thermal photon dephasing is now above 1 ms, even with an external microwave environment temperature of 115 mK.  Thus, the dissipator shows the capacity to support long-lived qubit coherence even in systems with elevated bath temperatures.

We note that the bath temperature of 115 mK is far above the state of the art, which is typically around 60 mK \cite{yehMicrowaveAttenuatorsUse2017}. This is likely due to our setup using microwave attenuators on the measurement feedline that are prone to poor thermalization, which was done deliberately so that thermal photon dephasing would be visible above other dephasing processes. We used a well-thermalized attenuator for the last 20 dB of attenuation on the loss line, and so the dissipator's bath temperature (and thus the final cavity temperature under refrigeration) is likely lower. In a real implementation, the loss line would be terminated; microwave terminations can be formed by encasing an antenna in a lossy epoxy, providing an excellent thermalization geometry \cite{wangCavityAttenuatorsSuperconducting2019}. We thus expect a dissipator implemented with a state-of-the-art attenuator or termination to have a bath temperature at or below $60$ mK, and thus to be able to cool the cavity below state of the art temperatures. Further gains could be made by raising the filter frequency; a small increase to 10 GHz would reduce the excitation number to $3.3\times10^{-4}$ at 60 mK. While such low excitation numbers are unnecessary for preventing thermal photon dephasing (due to the dissipator's damping effect described above), they raise the possibility of using the dissipator for rapid and high-fidelity reset of a qubit, which we plan to pursue in future work.

In order to be used for continuous cavity refrigeration, it is essential that the dissipation we drive is frequency-selective, otherwise it would cause unwanted damping of the qubit. To test this, we perform measurements of the qubit lifetime $T_1$ with and without a cavity refrigeration drive. We have measured up to the strongest dissipation drives possible with our setup---i.e., the strongest drives shown in Fig.~\ref{fig:ringdown} with $g_\mathrm{p}/2\pi \approx 10$ MHz and $\gamma_\mathrm{cav} = \kappa_\mathrm{eff} \approx 60\ \mu\mathrm{s}^{-1}$---and see no change in the mean $\bar{T_1} = 27\ \mu\mathrm{s}$.

\section{Conclusion}
We have presented a simple modular dissipator element that can be used to induce tunable, on-demand dissipation in a target mode. The dissipator is based on a driven parametric coupling between the target and a lossy mode, with the loss coming from coupling to a terminated microwave feedline. We have demonstrated how our dissipator can be used with a standard circuit QED readout cavity as its target. We can rapidly reset the cavity after measurement and cool it below its equilibrium temperature, without detectably affecting readout fidelity (at the $81 \%$ level) and without reducing qubit lifetime (at the $27\ \mu\mathrm{s}$ level) or coherence. In the case where qubit coherence is limited by background cavity photon population, our dissipator can be used to suppress dephasing and prolong qubit coherence. 

To add the dissipator to a standard cQED system, three components are required: (1) the dissipator itself (a tunable transmon coupler), (2) a high-frequency filter mode, and (3) a fast flux line. However, we note that parametric couplers similar to the dissipator are already present in many superconducting quantum processor architectures. The dissipator could therefore be integrated into these architectures with minimal extra complexity, as the coupler and flux line would be pre-existing; all that is required is to add an additional coupling to a high-frequency filter mode and to the target system (e.g., a readout cavity). In order to prevent unwanted loss on other modes (e.g., qubits that use this coupler for gates), the filter could be designed for strong frequency selectivity \cite{zhouHighsuppressionratioWideBandwidth2024}; alternately, to reduce complexity, a higher mode of a pre-existing broadband Purcell filter could be used. It should also be possible to multiplex the dissipator, using a single dissipator to couple to multiple targets and sequentially or simultaneously driving swaps between them and a lossy filter mode. In the case of simultaneous drives, care would need to be taken to remain in the overdamped regime so that excitations were not swapped between targets; as we have shown, it is possible to achieve excellent performance in this regime. Multiplexing is especially important, as  multiplexed cavities sharing a readout line can see elevated temperatures when one cavity is driven, due to heating of attenuators and other lossy components. Active cooling with a dissipator should alleviate this effect.

Future work could further optimize the dissipator for use in practical quantum processors. The filter frequency could be raised to further lower the cavity temperature, and its bandwidth could be reduced to allow similar dissipation strength with weaker driving. The off-chip termination could instead be implemented as a thin-film resistor either on chip or on the microwave launch, saving space. Driven dissipation is fully compatible with pulse-shaping techniques for rapid ringdown, and the two could be combined for even faster cavity reset. Cavity cooling could be used to preserve coherence even with reduced base-temperature attenuation and reduced isolation from following amplifiers, removing bulk and heat load from cryogenic setups. We note that cavity cooling also allows dispersive shift, cavity linewdith, and cavity frequency to be optimized for readout fidelity, without having to make tradeoffs to avoid photon-induced dephasing.

Future work can also use the dissipator as a source of loss for other applications, especially unconditional qubit reset. In a geometry like ours, this reset could be achieved by swapping qubit excitations into a lossy dissipator (which couples to the qubit via the cavity) or by driving a low-loss dissipator (i.e., a parametric coupler) with two tones to swap qubit excitations into a mode of the readout cavity. The latter approach would also work if the dissipator/coupler was instead coupled to the qubit, as in standard tunable coupler geometries. Such reset is useful in standard architectures, in bosonic encoding error suppression schemes \cite{gertlerProtectingBosonicQubit2021,kudraExperimentalRealizationDeterministic2022, lachance-quirionAutonomousQuantumError2023}, and generation and stabilization of many-body states \cite{miStableQuantumCorrelatedMany2023,wangDissipativePreparationStabilization2023}. We note that the dissipator's refrigeration capability should enable high fidelity qubit state preparation for qubits at all frequencies. Slight modifications could allow the dissipator to be used to drive nonreciprocal interactions \cite{metelmannNonreciprocalSignalRouting2018} or stabilize multi-qubit entangled states \cite{maStabilizingBellStates2019}. Finally, we note that the dissipator could allow researchers to study the fundamental behavior of quantum systems under lossy interactions, as the loss can be made strong for a short period and then effectively turned off during measurement.

\begin{acknowledgments}
The authors gratefully acknowledge Archana Kamal, Daniel Lidar, and Yao Lu for useful discussions. The TWPA amplifier used for these measurements was provided by MIT Lincoln Laboratory. Some devices used in development were fabricated and provided by the Superconducting Qubits at Lincoln Laboratory (SQUILL) Foundry at MIT Lincoln Laboratory, with funding from the Laboratory for Physical Sciences (LPS) Qubit Collaboratory.. HZ, VM, DK, AK, DMH, CM, JL, SS, AZ, and EMLF acknowledge funding from the National Science Foundation (NSF) under Grant No. OMA-1936388, the Office of Naval Research (ONR) under Grant No. N00014-21-1-2688, Research Corporation for Science Advancement under Cottrell Award 27550, and the Defense Advanced Research Projects Agency (DARPA) under MeasQUIT HR0011-24-9-0362. KWM and DK acknowledge support from NSF Grant No. PHY-1752844 (CAREER) and the Air Force Office of Scientific Research (AFOSR) Multidisciplinary University Research Initiative (MURI) Award on Programmable systems with non-Hermitian quantum dynamics (Grant No. FA9550- 21-1- 0202). 
\end{acknowledgments}

\appendix
\renewcommand{\thefigure}{S\arabic{figure}}
\renewcommand{\thetable}{S\Roman{table}}

\section{Parametric Exchange Interaction}
\label{app:parametric}
We derive the dissipator exchange with the cavity under the static Jaynes-Cummings Hamiltonian given in Eq.~\ref{eq:hamiltonian} and parametric drive given in Eq.~\ref{eq:drive}. Again we define the cavity-dissipator detuning $\Delta \equiv \omega_\mathrm{c} - \omega_\mathrm{diss}$. If the mean dissipator frequency is different under driving than the undriven frequency, as is the case for a flux drive on a SQUID, we simply replace $\omega_\mathrm{diss}$ by the mean driven frequency $\bar{\omega}_\mathrm{diss}$.
In the dispersive regime where $g_\mathrm{c}\ll \Delta$, and when the drive amplitude is small $\epsilon_\mathrm{p}\ll\Delta$, we can derive an analytic expression for the Rabi rate between the cavity-qubit system by treating the drive as a time-dependent perturbation to the Hamiltonian in Eq.~\ref{eq:hamiltonian}. We denote the bare dissipator-cavity eigenbasis as $|g,0\rangle$, $|e,0\rangle$, $|g,1\rangle$, $|e,1\rangle$,$\dots$,$|g,n\rangle$, $|e,n\rangle$, as shown in the energy level diagram Fig.~\ref{fig:circuit_diagram}(a). Since we are interested in the transition $|g,n+1\rangle \leftrightarrow |e,n\rangle$, indicated by blue arrows, we evaluate its Rabi frequency:
\begin{align}
	\Omega_R = \frac{1}{2} \sqrt{(\omega_\mathrm{p}-\Delta)^2+|\delta\omega_{g,n+1\rightarrow e,n}|^2},
\end{align}
where $\delta\omega$ denotes the transition amplitude between the two states. Treating the parametric drive as a perturbation to the Hamiltonian in Eq.~\ref{eq:hamiltonian}, the transition amplitude $\delta\omega$ can be calculated as
\begin{align}\label{eq:matrixElement}
\delta\omega_{g,n+1\rightarrow e,n} = \langle \psi^\prime_{g,n+1}|V_p|\psi^\prime_{e,n}\rangle 
	=-\frac{2\sqrt{n+1}g_\mathrm{c}\epsilon_\mathrm{p}}{\Delta},
\end{align}
where $V_p\equiv \epsilon_\mathrm{p} \sigma_z=\epsilon_\mathrm{p} \left(|g\rangle\langle g|-|e\rangle\langle e|\right)$ is the time-independent part of the parametric drive $H_p(t)$, and 
\begin{align}\label{eq:states}
\psi^\prime_{e,n}&=|e,n\rangle - \sqrt{n+1}\frac{g_c}{\Delta}\ket{g,n+1} \\
\psi^\prime_{g,n+1}&=|g,n+1\rangle+\sqrt{n+1}\frac{g_c}{\Delta}|e,n\rangle
\end{align}
are (up to a normalization) an eigenbasis of the Hamiltonian in Eq.~\ref{eq:hamiltonian}, obtained by treating the static cavity-dissipator interaction as a first-order perturbation to the uncoupled system. 
When the drive is on resonance, i.e., $\omega_\mathrm{p}=\Delta$, the Rabi frequency is given by:
\begin{align}\label{eq:swap_rate_app}
	\Omega_R = \frac{\delta\omega_{g,n+1\rightarrow e,n}}{2} = \frac{\sqrt{n+1}g_\mathrm{c}\epsilon_\mathrm{p}}{\Delta}.
\end{align}
The transition probability of $|g,n+1\rangle \rightarrow |e,n\rangle$ follows:
\begin{align}\label{eq:rabi_prob}
	P(|n+1,g\rangle \leftrightarrow |n,e\rangle) = \left(\frac{g_\mathrm{c}\epsilon_\mathrm{p}}{\Delta\Omega_R}\right)^2 \sin^2\left(\Omega_Rt\right).
\end{align}
Eq.~\eqref{eq:rabi_prob} shows that on resonance a full population inversion between $|g,n+1\rangle$ and $|e,n\rangle$ occurs with at frequency $2\Omega_R$, which is first order in $g_\mathrm{c}/\Delta$.

The parametric swap rate can also be derived by moving to a rotating frame defined by
\begin{align}\label{eq:Ut}
	U(t) = \exp\left(-i\omega_\mathrm{c}a^\dagger a t\right)\exp\left(-i\int_o^t\omega_\mathrm{p}(t^\prime)dt^\prime \frac{\sigma_z}{2}\right) ,
\end{align}
where $\omega_\mathrm{p}(t)=\bar{\omega}_\mathrm{diss} + \epsilon_\mathrm{p} \sin\left(\omega_\mathrm{p}t\right)$ is the dissipator frequency under the drive modulation.
In this rotating frame, the effective Hamiltonian can be calculated from:
\begin{align}\label{eq:Heff}
	\tilde{H} = UHU^\dagger + iU^\dagger \dot{U},
\end{align}
where $\tilde{H}$ effectively generates the evolution for the state $\tilde{|\psi\rangle}=U(t)|\psi\rangle$ under the Schr\"{o}dinger equation $\tilde{H}\tilde{|\psi\rangle} = i\partial\tilde{|\psi\rangle}/\partial t $. 

Substituting Eq.~\eqref{eq:Ut} into Eq.~\eqref{eq:Heff}, we get
\begin{align}\label{eq:Heff_inter}
	\tilde{H} = g_\mathrm{c} Ua^\dagger \sigma_- U^\dagger + h.c.,
\end{align}
which allows us to evaluate the effective cavity-dissipator coupling in the rotating frame.
Denoting $\theta_\mathrm{c}\equiv\omega_\mathrm{c}t$, $\theta_\mathrm{p}(t) = \frac{1}{2}\int_0^t dt^\prime \omega_\mathrm{p}(t^\prime)$, one can show:
\begin{align}
	\exp\left(-i\theta_\mathrm{c} a^\dagger a\right) a^\dagger \exp\left(i\theta_\mathrm{c} a^\dagger a\right) = a^\dagger \exp\left(i\theta_\mathrm{c}\right).
\end{align}
Similarly,
\begin{align}
	\exp\left(-i\theta_p\sigma_z\right)\sigma_-\exp\left(i\theta_p\sigma_z\right)=\sigma_- e^{-2i\theta_p}
\end{align}
Integrating for $\theta_p(t)$, Eq.~\eqref{eq:Heff_inter} becomes:
\begin{align}
	\tilde{H}=g_\mathrm{c}a^\dagger \sigma_- e^{-\frac{i\epsilon_\mathrm{p}}{\omega_\mathrm{p}}\cos\omega_\mathrm{p}t} + h.c..
\end{align}
Using the Jacobi-Anger expansion below:
\begin{align}
	e^{i z \cos \theta} \equiv \sum_{n=-\infty}^{\infty} i^n J_n(z) e^{i n \theta}
\end{align}
where $J_n(z)$ are Bessel functions of the first kind, we arrive at the final expression:
\begin{align}\label{eq:Heff_final}
	\tilde{H} = &g_\mathrm{c} a^\dagger\sigma_-J_0\left(\frac{\epsilon_\mathrm{p}}{\omega_\mathrm{p}}\right)e^{-i\Delta t } + h.c. \notag \\ 
	& +g_\mathrm{c} a^\dagger\sigma_-\sum_1^\infty (-i)^nJ_n\left(\frac{\epsilon_\mathrm{p}}{\omega_\mathrm{p}}\right)e^{-i\left(n\omega_\mathrm{p}-\Delta\right)t} + h.c..
\end{align}. 

As shown in Eq.~\eqref{eq:Heff_final}, the $n$-th sideband transition can be turned on if we choose a drive frequency at $\omega_\mathrm{p} = \Delta/n$; the effective qubit-cavity coupling strength is given by:
\begin{align}
	g_n = g_\mathrm{c} J_n(\frac{\epsilon_\mathrm{p}}{\omega_\mathrm{p}}),
\end{align}
which corresponds to an $n$-photon transition process. In the limit of small driving, when the first sideband is resonant with the cavity ($\omega_\mathrm{p} = \Delta$) this reduces to
\begin{align}
    g_p = g_1 \approx \frac{g_\mathrm{c} \epsilon_\mathrm{p}}{\Delta}
\end{align}
in agreement with Eq.~\ref{eq:swap_rate_app}. The analysis in the rotating frame shows that the sideband swap rate under the drive is of the first order in $g_\mathrm{c}/\Delta$ in the dispersive regime, confirming the result from the perturbation analysis.

\section{Driven Refrigeration}
\label{app:refrigeration}
Here we derive the amount of driven refrigeration in the general case. At equilibrium, the \emph{undriven} photon population of the cavity $\bar{n}_\mathrm{0}$ is given by the detailed balance of photon subtraction and addition due to the environmental photon loss and addition rates $\gamma_\mathrm{c}^\pm$ :
\begin{equation}
\label{eq:undrivenBalance}
    \bar{n}_\mathrm{0} = \frac{\gamma_\mathrm{c}^+}{\gamma_\mathrm{c}^-} .
\end{equation}
When the drive is turned on, photon loss and addition can also occur via the dissipator coupling with rate $\kappa_\mathrm{eff}$ (in the underdamped case, $\kappa_\mathrm{eff} = g_\mathrm{p}$) and the cavity photon population is
\begin{align}
    \label{eq:drivenBalanceCav}
    \bar{n}_\mathrm{c} = \frac{\gamma_\mathrm{c}^+ + \kappa_\mathrm{eff} \bar{n}_\mathrm{diss}}{\gamma_\mathrm{c}^- + \kappa_\mathrm{eff}}
\end{align} 
where $\bar{n}_\mathrm{diss}$ is the photon population of the dissipator. There is a similar equation for the dissipator:
\begin{align}
    \label{eq:drivenBalanceDiss}
    \bar{n}_\mathrm{diss} = \frac{\gamma_\mathrm{diss}^+ + \kappa_\mathrm{eff} \bar{n}_\mathrm{c}}{\gamma_\mathrm{diss}^- + \kappa_\mathrm{eff}}
\end{align} 
We can substitute Eq.~\ref{eq:drivenBalanceDiss} into Eq.~\ref{eq:drivenBalanceCav} and solve for $\bar{n}_\mathrm{c}$:
\begin{align}
    \label{eq:drivenBalance}
    \bar{n}_\mathrm{c} = \frac{\gamma_\mathrm{c}^+ \gamma_\mathrm{diss}^- + \kappa_\mathrm{eff} (\gamma_\mathrm{diss}^+ + \gamma_\mathrm{c}^+)}{\gamma_\mathrm{c}^-\gamma_\mathrm{diss}^- + \kappa_\mathrm{eff} (\gamma_\mathrm{c}^- + \gamma_\mathrm{diss}^-)}
\end{align} 
We can then subtract the undriven photon number $\bar{n}_0$ from Eq.~\ref{eq:drivenBalanceCav} to find
\begin{align}
    \label{eq:photonDifference}
    \bar{n}_\mathrm{c}-\bar{n}_\mathrm{0} = \frac{\kappa_\mathrm{eff}(\gamma_\mathrm{c}^-\gamma_\mathrm{diss}^+ - \gamma_\mathrm{c}^+\gamma_\mathrm{diss}^-)}{\gamma_\mathrm{c}^-[\gamma_\mathrm{diss}^- \kappa_\mathrm{eff} + \gamma_\mathrm{c}^+(\gamma_\mathrm{diss}^- + \kappa_\mathrm{eff})]}
\end{align}
When this is less than 0, driving removes photons from the cavity on average and thus cools it. This holds so long as $\gamma_\mathrm{c}^-\gamma_\mathrm{diss}^+ < \gamma_\mathrm{c}^+ \gamma_\mathrm{diss}^-$, i.e., as long as the \emph{undriven} dissipator population $\gamma_\mathrm{diss}^+ / \gamma_\mathrm{diss}^-$ is less than the undriven cavity population $\gamma_\mathrm{c}^+ / \gamma_\mathrm{c}^-$. In the case where the populations are thermal, this is simply the statement that $\omega_\mathrm{diss}/T_\mathrm{bath} > \omega_\mathrm{c} / T_0$, the condition given in Section \ref{sec:refTheory}.

\section{Fabrication Details}
\label{app:fab}
Devices are fabricated on chips diced from a [100] intrinsic silicon wafer with a resistivity $> 10,000~\mathrm{k\Omega \cdot cm}$. The chip is first cleaned using a Piranha solution consisting of sulfuric acid and hydrogen peroxide heated to $120^{\circ}\mathrm{C}$ for $10$ minutes to remove the organic contaminants, followed by a soak in a buffered oxide etch (BOE) solution with a $6{:}1$ concentration (6 parts by volume 40\% ammonium fluoride and 1 part by volume 49\% HF) for $5$ minutes to remove the native oxide layer on the wafer. The chip is then coated with a bi-layer stack of electron-beam resists (MMA EL13, PMMA A6). We utilize a $100~\mathrm{kV}$ Raith electron-beam lithography tool to realize the junctions as well as the bulk structure of the device. Note that for the small structures (${\sim}100~\mathrm{nm}$), we use a low beam current of $200~\mathrm{pA}$ to achieve fine resolution. Immediately after finishing the lithography, the sample is developed in a MIBK/IPA 1{:}3 solution at $0^{\circ}\mathrm{C}$ and then ashed in an oxygen plasma for $30$ seconds at $60~\mathrm{W}$ to remove all the remaining resist residue. To minimize the dielectric loss, the developed mask is again cleaned using the BOE solution for $30$~seconds and pumped down inside an Angstrom Engineering electron-beam evaporator within $5$ minutes of cleaning. Josephson junctions are deposited by employing the ``Manhattan Style'' evaporation scheme \cite{pottsCMOSCompatibleFabrication2001,costacheLateralMetallicDevices2012} with two evaporations at $40$ degree tilt separated by a $90$ degree azimuthal rotation and an oxidation step for $12$ minutes at $10$ Torr. Liftoff takes place in acetone heated to $45^{\circ}\mathrm{C}$ for $3$ hours followed by a sonication step in IPA, and methanol to clean the surface of the sample. At last, the device is post-ashed for one minute at $60~\mathrm{W}$ and coated with a protective resist prior to final dicing. Devices are wire-bonded in gold-plated copper boxes after removing the protective resist layer in acetone, IPA, and methanol solutions. 

\section{Apparatus Details}
\label{app:apparatus}
See Figure \ref{fig:fridge diagram} for a schematic of our measurement setup. All pulses are generated as single-sideband tones by mixing a microwave local oscillator (LO) tone in a Marki Microwave IQ mixer with two intermediate-frequency (IF) pulses from a Quantum Machines OPX unit. Readout and cavity population pulses are fed through a variable attenuator before being combined with qubit drive pulses and fed into the main input line. Fast flux pulses are similarly generated and attenuated before being fed into a separate line and then combined at low temperature with DC flux bias via a bias tee. A traveling-wave parametric amplifier (TWPA) is pumped with a constant tone and is used as a first-stage amplifier for readout signals, which are then fed through a HEMT amplifier and room-temperature amplifiers before demodulation and digitization in the same OPX unit. All lines are heavily attenuated and filtered with K\&L 12 GHz reactive low-pass filters and/or Eccosorb absorbative infrared filters. The line connected to the filter port (port 4) of the device is heavily attenuated and terminated at room temperature. To measure transmission through the filter, this port and the normal output port (port 2) are connected to a vector network analyzer (VNA); the VNA is removed during all other measurements.

\begin{table*}[t!]

\begin{center}
\centering
\begin{tabular}{||c|c|c|c|c|c|c|c|c||} 
\hline 
& \thead{Frequency (GHz)} & \thead{Anharmonicity (MHz)} & \thead{$\chi/2\pi$ (kHz)} & \thead{$\kappa/2\pi$ (MHz)} 
& \thead{$g/2\pi$ (MHz)}
& \thead{$T_1\ (\mu\mathrm{s})$} &\thead{$T_2^*\ (\mu\mathrm{s})$} \\ [0.3ex] 
\hline\hline
\thead{Qubit} & 3.368 & -175 & 200 & - & 53.9 & 27 & 4 \\ [0.3ex] 
\hline
\thead{Cavity} & 5.594 & - & 200 & 0.477 & 145 & - & - \\ [0.3ex]
\hline
\thead{Dissipator} & 15.3 - 4.2 & -350 & - & - & - & $<0.05$ & $<0.1$ \\ [0.3ex] 
\hline
\thead{Filter} & 8.6 & - & - & 120 & 535 & - & - \\ [0.3ex] 
 \hline
 
\end{tabular}
\end{center}
\caption{Measured or inferred parameters of the device used in the experiment. All mode frequencies, qubit anharmonicity, cavity and filter linewidths, and qubit-cavity dispersive shift are directly measured spectroscopically. Qubit-cavity coupling $g_\mathrm{q}$ (in the qubit row) is inferred from measurements of qubit and cavity frequencies $\omega_\mathrm{q},\omega_\mathrm{c}$, qubit anharmonicity $\alpha_\mathrm{q}$, and qubit-cavity dispersive shift $\chi$. Dissipator anharmonicity is inferred from simulations of the dissipator capacitance. Dissipator-cavity coupling $g_\mathrm{c}$ (in the cavity row) and dissipator-filter coupling $g_\mathrm{f}$ (in the filter row) are fit from avoided crossings in spectroscopy. Qubit lifetime $T_1$ and Ramsey coherence time $T_2^*$ are directly measured; dissipator coherence is unknown, but can be bounded as shorter than $0.05\ \mu\mathrm{s}$ lifetime based on a calculation of Purcell decay.} 
\label{tab:sim}

\end{table*}

\section{Device Parameter Calibration}\label{app:calib}
Several methods are used to measure and infer the system parameters. The spectroscopic measurements reported in Fig.~\ref{fig:fluxtuning} are performed using a vector network analyzer to measure transmission from port 4 to port 2, and from port 1 to port 2, as a function of drive frequency. Features in the transmission spectrum directly give the mode frequencies of the cavity and filter $\omega_\mathrm{c}/2\pi = 5.594$ GHz, $\omega\mathrm{f}/2\pi = 8.6$ GHz and their linewidths $\kappa_\mathrm{c}/2\pi = 620$ kHz, $\kappa_\mathrm{f}/2\pi \approx 120$ MHz. The filter linewidth is approximate as the broad line is highly non-Lorentzian. By fitting the locations of avoided crossings between dissipator and cavity and filter modes, as well as other spurious modes on chip, we can fit the dissipator frequency to the general form
\begin{align}
\omega_\mathrm{diss}(\phi) = &(\omega_\mathrm{diss,max} - \alpha_\mathrm{diss}) (\cos^2\pi\phi+d^2*\sin^2\pi\phi)^{1/4}  \notag \\
+ &\alpha_\mathrm{diss}
\end{align}
where $\phi$ is the flux in units of flux quanta, $d \equiv (E_{J2}-E_{J1})/(E_{J2}+E_{J1})$ captures any junction asymmetry, and $\alpha_\mathrm{diss}/2\pi\approx -350$ MHz is the dissipator anharmonicity, which we extract from finite-element simulations of its capacitance. We extract $\omega_\mathrm{diss,max}/2\pi = 15.3$ GHz and $d = 0.085$, leading to a dissipator frequency which tunes from 15.3 GHz down to 4.2 GHz. We fit the avoided crossings themselves to extract the dissipator-filter coupling $g_\mathrm{f}/2\pi = 535$ MHz and the dissipator-cavity coupling $g_\mathrm{c}^\prime/2\pi = 118$ MHz at the avoided crossing. When we tune the dissipator into resonance with the filter, its Josephson energy rises by a factor of 2.25, and so the coupling strength rises by a factor of $2.25^{1/4} = 1.23$ to $g_\mathrm{c}/2\pi = 145$ MHz \cite{kochChargeinsensitiveQubitDesign2007}. We note that this value (and thus any calculated $g_\mathrm{p}$ value) is approximate, as the mode structure may change as the dissipator tunes through frequencies.  

We take pulsed spectroscopy measurements of the qubit to find its frequency $\omega_q$ and anharmonicity $\alpha_{q}$. We perform spectroscopy of the cavity resonance with and without a population-inverting $\pi$ pulse on the qubit to measure the dispersive shift $\chi$, and from it extract the qubit-cavity coupling $g_\mathrm{q}$. We measure mean qubit lifetime $\bar{T_1} = 27\ \mu\mathrm{s}$ and coherence $\bar{T_2^*} = 4\ \mu\mathrm{s}$ with ordinary decay and Ramsey experiments. Similar measurements on the dissipator yield no signal, which is expected as the dissipator lifetime should be Purcell-limited to below 50 ns at all dissipator frequencies. All device parameters are shown in Table \ref{tab:sim}.

We measure the qubit's excited state population by driving Rabi oscillations on the $\ket{e}\leftrightarrow \ket{f}$ transition frequency with and without a $\ket{g}\leftrightarrow \ket{e}$ pi pulse beforehand. Taking the ratio of the amplitudes of these two signals gives us the ratio of the $\ket{g}$ and $\ket{e}$ state populations, and thus the qubit temperature \cite{jinThermalResidualExcitedState2015}. We find an excited state occupation of 3.4\% and thus a temperature of 47 mK, typical for transmon qubits in this frequency range. We expect that this temperature is due to the detailed balance between relaxation due to coupling to cold two-level systems, and excitation due to hot quasiparticles \cite{serniakHotNonequilibriumQuasiparticles2018a}.

\begin{figure*}
    \centering
    \includegraphics{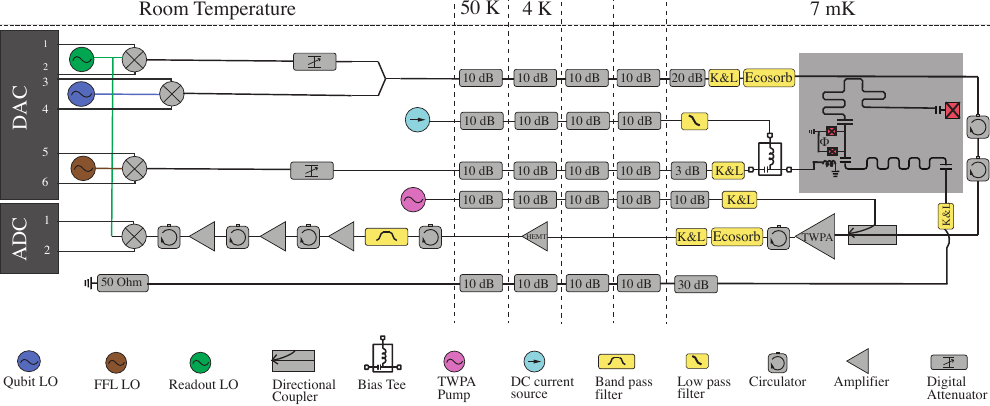}
    \caption{\label{fig:fridge diagram}
    Experimental schematic. Qubit and cavity drives are generated with dual-channel baseband pulses fed into IQ mixers and mixed with local oscillators (LOs) to create single sideband microwave pulses. The cavity drives pass through a variable attenuator before being combined with qubit drives and fed into the fridge on a heavily attenuated and filtered input line. At base temperature the lines are filtered with Eccosorb infrared filters from Quantum Microwave and 12 GHz low-pass filters from K\&L. After transmitting through the device, readout signals pass through terminated circulators (i.e., isolators) before being amplified by a TWPA; the TWPA pump is generated separately at room temperature and fed in on its own attenuated and filtered line before being coupled into the amplifier. The amplified signals are further isolated before passing through another stage of filtering and traveling back up the fridge, through a HEMT semiconductor amplifier at $\sim 4$ K and successive stages of amplifiers at room temperature, with isolation between each stage. The signal is finally demodulated with the same LO used to generate the readout pulses, before being digitized and further demodulated in software. Microwave fast flux pulses are generated by dual-channel baseband pulses mixed with another LO, passed through a variable attenuator, and fed into the fridge through an attenuated and 12 GHz low-pass filtered line. Static flux bias travels down an attenuated line with a 250 MHz low-pass filter at base temperature, and is combined with the microwave flux pulses with a bias tee at base temperature before being fed into the device. The lossy ``termination'' line connects to the device via heavy attenuation and another K\&L filter. Not shown in this configuration, sometimes a vector network analyzer (VNA) is connected to the termination line and the ordinary output line in order to measure transmission through the filter.
    }
\end{figure*}

To calibrate the flux drive through the dissipator SQUID, we first tune the dissipator through a full flux quantum using DC injected on the fast flux line. We thus obtain a current-to-flux transfer function. We then measure the microwave power in our parametric drive at the input to the fridge, and use the same transfer function to transform it to a flux drive amplitude. We include an extra 3 dB attenuation to account for the extra fixed attenuation in the microwave flux drive line (see Fig.~\ref{fig:fridge diagram}). We do not know the exact attenuation from the coaxial cables in the fridge, but we assume it is small at these frequencies (below 3.5 GHz) and so ignore it. Once we have the parametric drive amplitude in flux units, we convert this to frequency units (i.e., to $\epsilon_\mathrm{p}$) by taking the slope of the dissipator flux tuning curve fit above. We note that the curve is not truly linear, but in the small modulation limit that we use, the linear approximation is correct to a part in $10^4$ and so is fully appropriate. Finally, we use Eq.~\ref{eq:gp} to convert from this parametric drive amplitude to a parametric coupling rate $g_\mathrm{p}$. We note that this procedure is vulnerable to errors in our calibration of $g_\mathrm{c}$ as well as the flux drive amplitude, and so the values of $g_\mathrm{p}$ extracted should be viewed as approximate. One could instead use the induced cavity loss rate $\kappa_\mathrm{eff}$ to calibrate $g_\mathrm{p}$, but this is ``assuming the answer'' when it comes to the behavior of the driven dissipation. We therefore use our approximate calibration, but note that it closely matches the theoretical prediction.

\bibliography{references}

\begin{thebibliography}{56}%
\makeatletter
\providecommand \@ifxundefined [1]{%
 \@ifx{#1\undefined}
}%
\providecommand \@ifnum [1]{%
 \ifnum #1\expandafter \@firstoftwo
 \else \expandafter \@secondoftwo
 \fi
}%
\providecommand \@ifx [1]{%
 \ifx #1\expandafter \@firstoftwo
 \else \expandafter \@secondoftwo
 \fi
}%
\providecommand \natexlab [1]{#1}%
\providecommand \enquote  [1]{``#1''}%
\providecommand \bibnamefont  [1]{#1}%
\providecommand \bibfnamefont [1]{#1}%
\providecommand \citenamefont [1]{#1}%
\providecommand \href@noop [0]{\@secondoftwo}%
\providecommand \href [0]{\begingroup \@sanitize@url \@href}%
\providecommand \@href[1]{\@@startlink{#1}\@@href}%
\providecommand \@@href[1]{\endgroup#1\@@endlink}%
\providecommand \@sanitize@url [0]{\catcode `\\12\catcode `\$12\catcode `\&12\catcode `\#12\catcode `\^12\catcode `\_12\catcode `\%12\relax}%
\providecommand \@@startlink[1]{}%
\providecommand \@@endlink[0]{}%
\providecommand \url  [0]{\begingroup\@sanitize@url \@url }%
\providecommand \@url [1]{\endgroup\@href {#1}{\urlprefix }}%
\providecommand \urlprefix  [0]{URL }%
\providecommand \Eprint [0]{\href }%
\providecommand \doibase [0]{https://doi.org/}%
\providecommand \selectlanguage [0]{\@gobble}%
\providecommand \bibinfo  [0]{\@secondoftwo}%
\providecommand \bibfield  [0]{\@secondoftwo}%
\providecommand \translation [1]{[#1]}%
\providecommand \BibitemOpen [0]{}%
\providecommand \bibitemStop [0]{}%
\providecommand \bibitemNoStop [0]{.\EOS\space}%
\providecommand \EOS [0]{\spacefactor3000\relax}%
\providecommand \BibitemShut  [1]{\csname bibitem#1\endcsname}%
\let\auto@bib@innerbib\@empty
\bibitem [{\citenamefont {Kjaergaard}\ \emph {et~al.}(2020)\citenamefont {Kjaergaard}, \citenamefont {Schwartz}, \citenamefont {Braum{\"u}ller}, \citenamefont {Krantz}, \citenamefont {Wang}, \citenamefont {Gustavsson},\ and\ \citenamefont {Oliver}}]{kjaergaardSuperconductingQubitsCurrent2020}%
  \BibitemOpen
  \bibfield  {author} {\bibinfo {author} {\bibfnamefont {M.}~\bibnamefont {Kjaergaard}}, \bibinfo {author} {\bibfnamefont {M.~E.}\ \bibnamefont {Schwartz}}, \bibinfo {author} {\bibfnamefont {J.}~\bibnamefont {Braum{\"u}ller}}, \bibinfo {author} {\bibfnamefont {P.}~\bibnamefont {Krantz}}, \bibinfo {author} {\bibfnamefont {J.~I.-J.}\ \bibnamefont {Wang}}, \bibinfo {author} {\bibfnamefont {S.}~\bibnamefont {Gustavsson}},\ and\ \bibinfo {author} {\bibfnamefont {W.~D.}\ \bibnamefont {Oliver}},\ }\bibfield  {title} {\bibinfo {title} {Superconducting {{Qubits}}: {{Current State}} of {{Play}}},\ }\href {https://doi.org/10.1146/annurev-conmatphys-031119-050605} {\bibfield  {journal} {\bibinfo  {journal} {Annual Review of Condensed Matter Physics}\ }\textbf {\bibinfo {volume} {11}},\ \bibinfo {pages} {369} (\bibinfo {year} {2020})}\BibitemShut {NoStop}%
\bibitem [{\citenamefont {Paik}\ \emph {et~al.}(2011)\citenamefont {Paik}, \citenamefont {Schuster}, \citenamefont {Bishop}, \citenamefont {Kirchmair}, \citenamefont {Catelani}, \citenamefont {Sears}, \citenamefont {Johnson}, \citenamefont {Reagor}, \citenamefont {Frunzio}, \citenamefont {Glazman}, \citenamefont {Girvin}, \citenamefont {Devoret},\ and\ \citenamefont {Schoelkopf}}]{paikObservationHighCoherence2011}%
  \BibitemOpen
  \bibfield  {author} {\bibinfo {author} {\bibfnamefont {H.}~\bibnamefont {Paik}}, \bibinfo {author} {\bibfnamefont {D.~I.}\ \bibnamefont {Schuster}}, \bibinfo {author} {\bibfnamefont {L.~S.}\ \bibnamefont {Bishop}}, \bibinfo {author} {\bibfnamefont {G.}~\bibnamefont {Kirchmair}}, \bibinfo {author} {\bibfnamefont {G.}~\bibnamefont {Catelani}}, \bibinfo {author} {\bibfnamefont {A.~P.}\ \bibnamefont {Sears}}, \bibinfo {author} {\bibfnamefont {B.~R.}\ \bibnamefont {Johnson}}, \bibinfo {author} {\bibfnamefont {M.~J.}\ \bibnamefont {Reagor}}, \bibinfo {author} {\bibfnamefont {L.}~\bibnamefont {Frunzio}}, \bibinfo {author} {\bibfnamefont {L.~I.}\ \bibnamefont {Glazman}}, \bibinfo {author} {\bibfnamefont {S.~M.}\ \bibnamefont {Girvin}}, \bibinfo {author} {\bibfnamefont {M.~H.}\ \bibnamefont {Devoret}},\ and\ \bibinfo {author} {\bibfnamefont {R.~J.}\ \bibnamefont {Schoelkopf}},\ }\bibfield  {title} {\bibinfo {title} {Observation of {{High Coherence}} in {{Josephson Junction Qubits Measured}} in a {{Three-Dimensional
  Circuit QED Architecture}}},\ }\href {https://doi.org/10.1103/PhysRevLett.107.240501} {\bibfield  {journal} {\bibinfo  {journal} {Physical Review Letters}\ }\textbf {\bibinfo {volume} {107}},\ \bibinfo {pages} {240501} (\bibinfo {year} {2011})}\BibitemShut {NoStop}%
\bibitem [{\citenamefont {Oliver}\ and\ \citenamefont {Welander}(2013)}]{oliverMaterialsSuperconductingQuantum2013}%
  \BibitemOpen
  \bibfield  {author} {\bibinfo {author} {\bibfnamefont {W.~D.}\ \bibnamefont {Oliver}}\ and\ \bibinfo {author} {\bibfnamefont {P.~B.}\ \bibnamefont {Welander}},\ }\bibfield  {title} {\bibinfo {title} {Materials in superconducting quantum bits},\ }\href {https://doi.org/10.1557/mrs.2013.229} {\bibfield  {journal} {\bibinfo  {journal} {MRS Bulletin}\ }\textbf {\bibinfo {volume} {38}},\ \bibinfo {pages} {816} (\bibinfo {year} {2013})}\BibitemShut {NoStop}%
\bibitem [{\citenamefont {Barends}\ \emph {et~al.}(2011)\citenamefont {Barends}, \citenamefont {Wenner}, \citenamefont {Lenander}, \citenamefont {Chen}, \citenamefont {Bialczak}, \citenamefont {Kelly}, \citenamefont {Lucero}, \citenamefont {O'Malley}, \citenamefont {Mariantoni}, \citenamefont {Sank}, \citenamefont {Wang}, \citenamefont {White}, \citenamefont {Yin}, \citenamefont {Zhao}, \citenamefont {Cleland}, \citenamefont {Martinis},\ and\ \citenamefont {Baselmans}}]{barendsMinimizingQuasiparticleGeneration2011}%
  \BibitemOpen
  \bibfield  {author} {\bibinfo {author} {\bibfnamefont {R.}~\bibnamefont {Barends}}, \bibinfo {author} {\bibfnamefont {J.}~\bibnamefont {Wenner}}, \bibinfo {author} {\bibfnamefont {M.}~\bibnamefont {Lenander}}, \bibinfo {author} {\bibfnamefont {Y.}~\bibnamefont {Chen}}, \bibinfo {author} {\bibfnamefont {R.~C.}\ \bibnamefont {Bialczak}}, \bibinfo {author} {\bibfnamefont {J.}~\bibnamefont {Kelly}}, \bibinfo {author} {\bibfnamefont {E.}~\bibnamefont {Lucero}}, \bibinfo {author} {\bibfnamefont {P.}~\bibnamefont {O'Malley}}, \bibinfo {author} {\bibfnamefont {M.}~\bibnamefont {Mariantoni}}, \bibinfo {author} {\bibfnamefont {D.}~\bibnamefont {Sank}}, \bibinfo {author} {\bibfnamefont {H.}~\bibnamefont {Wang}}, \bibinfo {author} {\bibfnamefont {T.~C.}\ \bibnamefont {White}}, \bibinfo {author} {\bibfnamefont {Y.}~\bibnamefont {Yin}}, \bibinfo {author} {\bibfnamefont {J.}~\bibnamefont {Zhao}}, \bibinfo {author} {\bibfnamefont {A.~N.}\ \bibnamefont {Cleland}}, \bibinfo {author} {\bibfnamefont {J.~M.}\ \bibnamefont
  {Martinis}},\ and\ \bibinfo {author} {\bibfnamefont {J.~J.~A.}\ \bibnamefont {Baselmans}},\ }\bibfield  {title} {\bibinfo {title} {Minimizing quasiparticle generation from stray infrared light in superconducting quantum circuits},\ }\href {https://doi.org/10.1063/1.3638063} {\bibfield  {journal} {\bibinfo  {journal} {Applied Physics Letters}\ }\textbf {\bibinfo {volume} {99}},\ \bibinfo {pages} {113507} (\bibinfo {year} {2011})}\BibitemShut {NoStop}%
\bibitem [{\citenamefont {Nguyen}\ \emph {et~al.}(2019)\citenamefont {Nguyen}, \citenamefont {Lin}, \citenamefont {Somoroff}, \citenamefont {Mencia}, \citenamefont {Grabon},\ and\ \citenamefont {Manucharyan}}]{nguyenHighCoherenceFluxoniumQubit2019}%
  \BibitemOpen
  \bibfield  {author} {\bibinfo {author} {\bibfnamefont {L.~B.}\ \bibnamefont {Nguyen}}, \bibinfo {author} {\bibfnamefont {Y.-H.}\ \bibnamefont {Lin}}, \bibinfo {author} {\bibfnamefont {A.}~\bibnamefont {Somoroff}}, \bibinfo {author} {\bibfnamefont {R.}~\bibnamefont {Mencia}}, \bibinfo {author} {\bibfnamefont {N.}~\bibnamefont {Grabon}},\ and\ \bibinfo {author} {\bibfnamefont {V.~E.}\ \bibnamefont {Manucharyan}},\ }\bibfield  {title} {\bibinfo {title} {High-{{Coherence Fluxonium Qubit}}},\ }\href {https://doi.org/10.1103/PhysRevX.9.041041} {\bibfield  {journal} {\bibinfo  {journal} {Physical Review X}\ }\textbf {\bibinfo {volume} {9}},\ \bibinfo {pages} {041041} (\bibinfo {year} {2019})}\BibitemShut {NoStop}%
\bibitem [{\citenamefont {Gyenis}\ \emph {et~al.}(2021)\citenamefont {Gyenis}, \citenamefont {Mundada}, \citenamefont {Di~Paolo}, \citenamefont {Hazard}, \citenamefont {You}, \citenamefont {Schuster}, \citenamefont {Koch}, \citenamefont {Blais},\ and\ \citenamefont {Houck}}]{gyenisExperimentalRealizationProtected2021}%
  \BibitemOpen
  \bibfield  {author} {\bibinfo {author} {\bibfnamefont {A.}~\bibnamefont {Gyenis}}, \bibinfo {author} {\bibfnamefont {P.~S.}\ \bibnamefont {Mundada}}, \bibinfo {author} {\bibfnamefont {A.}~\bibnamefont {Di~Paolo}}, \bibinfo {author} {\bibfnamefont {T.~M.}\ \bibnamefont {Hazard}}, \bibinfo {author} {\bibfnamefont {X.}~\bibnamefont {You}}, \bibinfo {author} {\bibfnamefont {D.~I.}\ \bibnamefont {Schuster}}, \bibinfo {author} {\bibfnamefont {J.}~\bibnamefont {Koch}}, \bibinfo {author} {\bibfnamefont {A.}~\bibnamefont {Blais}},\ and\ \bibinfo {author} {\bibfnamefont {A.~A.}\ \bibnamefont {Houck}},\ }\bibfield  {title} {\bibinfo {title} {Experimental {{Realization}} of a {{Protected Superconducting Circuit Derived}} from the 0--$\pi$ {{Qubit}}},\ }\href {https://doi.org/10.1103/PRXQuantum.2.010339} {\bibfield  {journal} {\bibinfo  {journal} {PRX Quantum}\ }\textbf {\bibinfo {volume} {2}},\ \bibinfo {pages} {010339} (\bibinfo {year} {2021})}\BibitemShut {NoStop}%
\bibitem [{\citenamefont {Blais}\ \emph {et~al.}(2021)\citenamefont {Blais}, \citenamefont {Grimsmo}, \citenamefont {Girvin},\ and\ \citenamefont {Wallraff}}]{blaisCircuitQuantumElectrodynamics2021}%
  \BibitemOpen
  \bibfield  {author} {\bibinfo {author} {\bibfnamefont {A.}~\bibnamefont {Blais}}, \bibinfo {author} {\bibfnamefont {A.~L.}\ \bibnamefont {Grimsmo}}, \bibinfo {author} {\bibfnamefont {S.~M.}\ \bibnamefont {Girvin}},\ and\ \bibinfo {author} {\bibfnamefont {A.}~\bibnamefont {Wallraff}},\ }\bibfield  {title} {\bibinfo {title} {Circuit quantum electrodynamics},\ }\href {https://doi.org/10.1103/RevModPhys.93.025005} {\bibfield  {journal} {\bibinfo  {journal} {Reviews of Modern Physics}\ }\textbf {\bibinfo {volume} {93}},\ \bibinfo {pages} {025005} (\bibinfo {year} {2021})}\BibitemShut {NoStop}%
\bibitem [{\citenamefont {Gambetta}\ \emph {et~al.}(2006)\citenamefont {Gambetta}, \citenamefont {Blais}, \citenamefont {Schuster}, \citenamefont {Wallraff}, \citenamefont {Frunzio}, \citenamefont {Majer}, \citenamefont {Devoret}, \citenamefont {Girvin},\ and\ \citenamefont {Schoelkopf}}]{gambettaQubitphotonInteractionsCavity2006}%
  \BibitemOpen
  \bibfield  {author} {\bibinfo {author} {\bibfnamefont {J.}~\bibnamefont {Gambetta}}, \bibinfo {author} {\bibfnamefont {A.}~\bibnamefont {Blais}}, \bibinfo {author} {\bibfnamefont {D.~I.}\ \bibnamefont {Schuster}}, \bibinfo {author} {\bibfnamefont {A.}~\bibnamefont {Wallraff}}, \bibinfo {author} {\bibfnamefont {L.}~\bibnamefont {Frunzio}}, \bibinfo {author} {\bibfnamefont {J.}~\bibnamefont {Majer}}, \bibinfo {author} {\bibfnamefont {M.~H.}\ \bibnamefont {Devoret}}, \bibinfo {author} {\bibfnamefont {S.~M.}\ \bibnamefont {Girvin}},\ and\ \bibinfo {author} {\bibfnamefont {R.~J.}\ \bibnamefont {Schoelkopf}},\ }\bibfield  {title} {\bibinfo {title} {Qubit-photon interactions in a cavity: {{Measurement-induced}} dephasing and number splitting},\ }\href {https://doi.org/10.1103/PhysRevA.74.042318} {\bibfield  {journal} {\bibinfo  {journal} {Physical Review A}\ }\textbf {\bibinfo {volume} {74}},\ \bibinfo {pages} {042318} (\bibinfo {year} {2006})}\BibitemShut {NoStop}%
\bibitem [{\citenamefont {Clerk}\ and\ \citenamefont {Utami}(2007)}]{clerkUsingQubitMeasure2007a}%
  \BibitemOpen
  \bibfield  {author} {\bibinfo {author} {\bibfnamefont {A.~A.}\ \bibnamefont {Clerk}}\ and\ \bibinfo {author} {\bibfnamefont {D.~W.}\ \bibnamefont {Utami}},\ }\bibfield  {title} {\bibinfo {title} {Using a qubit to measure photon-number statistics of a driven thermal oscillator},\ }\href {https://doi.org/10.1103/PhysRevA.75.042302} {\bibfield  {journal} {\bibinfo  {journal} {Physical Review A}\ }\textbf {\bibinfo {volume} {75}},\ \bibinfo {pages} {042302} (\bibinfo {year} {2007})}\BibitemShut {NoStop}%
\bibitem [{\citenamefont {Sears}\ \emph {et~al.}(2012)\citenamefont {Sears}, \citenamefont {Petrenko}, \citenamefont {Catelani}, \citenamefont {Sun}, \citenamefont {Paik}, \citenamefont {Kirchmair}, \citenamefont {Frunzio}, \citenamefont {Glazman}, \citenamefont {Girvin},\ and\ \citenamefont {Schoelkopf}}]{photon-shot-dephasing-schoelkopf}%
  \BibitemOpen
  \bibfield  {author} {\bibinfo {author} {\bibfnamefont {A.~P.}\ \bibnamefont {Sears}}, \bibinfo {author} {\bibfnamefont {A.}~\bibnamefont {Petrenko}}, \bibinfo {author} {\bibfnamefont {G.}~\bibnamefont {Catelani}}, \bibinfo {author} {\bibfnamefont {L.}~\bibnamefont {Sun}}, \bibinfo {author} {\bibfnamefont {H.}~\bibnamefont {Paik}}, \bibinfo {author} {\bibfnamefont {G.}~\bibnamefont {Kirchmair}}, \bibinfo {author} {\bibfnamefont {L.}~\bibnamefont {Frunzio}}, \bibinfo {author} {\bibfnamefont {L.~I.}\ \bibnamefont {Glazman}}, \bibinfo {author} {\bibfnamefont {S.~M.}\ \bibnamefont {Girvin}},\ and\ \bibinfo {author} {\bibfnamefont {R.~J.}\ \bibnamefont {Schoelkopf}},\ }\bibfield  {title} {\bibinfo {title} {Photon shot noise dephasing in the strong-dispersive limit of circuit qed},\ }\href {https://doi.org/10.1103/PhysRevB.86.180504} {\bibfield  {journal} {\bibinfo  {journal} {Phys. Rev. B}\ }\textbf {\bibinfo {volume} {86}},\ \bibinfo {pages} {180504} (\bibinfo {year} {2012})}\BibitemShut {NoStop}%
\bibitem [{\citenamefont {Yan}\ \emph {et~al.}(2018)\citenamefont {Yan}, \citenamefont {Campbell}, \citenamefont {Krantz}, \citenamefont {Kjaergaard}, \citenamefont {Kim}, \citenamefont {Yoder}, \citenamefont {Hover}, \citenamefont {Sears}, \citenamefont {Kerman}, \citenamefont {Orlando}, \citenamefont {Gustavsson},\ and\ \citenamefont {Oliver}}]{yanDistinguishingCoherentThermal2018}%
  \BibitemOpen
  \bibfield  {author} {\bibinfo {author} {\bibfnamefont {F.}~\bibnamefont {Yan}}, \bibinfo {author} {\bibfnamefont {D.}~\bibnamefont {Campbell}}, \bibinfo {author} {\bibfnamefont {P.}~\bibnamefont {Krantz}}, \bibinfo {author} {\bibfnamefont {M.}~\bibnamefont {Kjaergaard}}, \bibinfo {author} {\bibfnamefont {D.}~\bibnamefont {Kim}}, \bibinfo {author} {\bibfnamefont {J.~L.}\ \bibnamefont {Yoder}}, \bibinfo {author} {\bibfnamefont {D.}~\bibnamefont {Hover}}, \bibinfo {author} {\bibfnamefont {A.}~\bibnamefont {Sears}}, \bibinfo {author} {\bibfnamefont {A.~J.}\ \bibnamefont {Kerman}}, \bibinfo {author} {\bibfnamefont {T.~P.}\ \bibnamefont {Orlando}}, \bibinfo {author} {\bibfnamefont {S.}~\bibnamefont {Gustavsson}},\ and\ \bibinfo {author} {\bibfnamefont {W.~D.}\ \bibnamefont {Oliver}},\ }\bibfield  {title} {\bibinfo {title} {Distinguishing {{Coherent}} and {{Thermal Photon Noise}} in a {{Circuit Quantum Electrodynamical System}}},\ }\href {https://doi.org/10.1103/PhysRevLett.120.260504} {\bibfield  {journal}
  {\bibinfo  {journal} {Physical Review Letters}\ }\textbf {\bibinfo {volume} {120}},\ \bibinfo {pages} {260504} (\bibinfo {year} {2018})}\BibitemShut {NoStop}%
\bibitem [{\citenamefont {Yeh}\ \emph {et~al.}(2017)\citenamefont {Yeh}, \citenamefont {LeFebvre}, \citenamefont {Premaratne}, \citenamefont {Wellstood},\ and\ \citenamefont {Palmer}}]{yehMicrowaveAttenuatorsUse2017}%
  \BibitemOpen
  \bibfield  {author} {\bibinfo {author} {\bibfnamefont {J.-H.}\ \bibnamefont {Yeh}}, \bibinfo {author} {\bibfnamefont {J.}~\bibnamefont {LeFebvre}}, \bibinfo {author} {\bibfnamefont {S.}~\bibnamefont {Premaratne}}, \bibinfo {author} {\bibfnamefont {F.~C.}\ \bibnamefont {Wellstood}},\ and\ \bibinfo {author} {\bibfnamefont {B.~S.}\ \bibnamefont {Palmer}},\ }\bibfield  {title} {\bibinfo {title} {Microwave attenuators for use with quantum devices below 100 {{mK}}},\ }\href {https://doi.org/10.1063/1.4984894} {\bibfield  {journal} {\bibinfo  {journal} {Journal of Applied Physics}\ }\textbf {\bibinfo {volume} {121}},\ \bibinfo {pages} {224501} (\bibinfo {year} {2017})}\BibitemShut {NoStop}%
\bibitem [{\citenamefont {Wang}\ \emph {et~al.}(2019)\citenamefont {Wang}, \citenamefont {Shankar}, \citenamefont {Minev}, \citenamefont {{Campagne-Ibarcq}}, \citenamefont {Narla},\ and\ \citenamefont {Devoret}}]{wangCavityAttenuatorsSuperconducting2019}%
  \BibitemOpen
  \bibfield  {author} {\bibinfo {author} {\bibfnamefont {Z.}~\bibnamefont {Wang}}, \bibinfo {author} {\bibfnamefont {S.}~\bibnamefont {Shankar}}, \bibinfo {author} {\bibfnamefont {Z.}~\bibnamefont {Minev}}, \bibinfo {author} {\bibfnamefont {P.}~\bibnamefont {{Campagne-Ibarcq}}}, \bibinfo {author} {\bibfnamefont {A.}~\bibnamefont {Narla}},\ and\ \bibinfo {author} {\bibfnamefont {M.}~\bibnamefont {Devoret}},\ }\bibfield  {title} {\bibinfo {title} {Cavity {{Attenuators}} for {{Superconducting Qubits}}},\ }\href {https://doi.org/10.1103/PhysRevApplied.11.014031} {\bibfield  {journal} {\bibinfo  {journal} {Physical Review Applied}\ }\textbf {\bibinfo {volume} {11}},\ \bibinfo {pages} {014031} (\bibinfo {year} {2019})}\BibitemShut {NoStop}%
\bibitem [{\citenamefont {McClure}\ \emph {et~al.}(2016)\citenamefont {McClure}, \citenamefont {Paik}, \citenamefont {Bishop}, \citenamefont {Steffen}, \citenamefont {Chow},\ and\ \citenamefont {Gambetta}}]{mcclureRapidDrivenReset2016}%
  \BibitemOpen
  \bibfield  {author} {\bibinfo {author} {\bibfnamefont {D.~T.}\ \bibnamefont {McClure}}, \bibinfo {author} {\bibfnamefont {H.}~\bibnamefont {Paik}}, \bibinfo {author} {\bibfnamefont {L.~S.}\ \bibnamefont {Bishop}}, \bibinfo {author} {\bibfnamefont {M.}~\bibnamefont {Steffen}}, \bibinfo {author} {\bibfnamefont {J.~M.}\ \bibnamefont {Chow}},\ and\ \bibinfo {author} {\bibfnamefont {J.~M.}\ \bibnamefont {Gambetta}},\ }\bibfield  {title} {\bibinfo {title} {Rapid {{Driven Reset}} of a {{Qubit Readout Resonator}}},\ }\href {https://doi.org/10.1103/PhysRevApplied.5.011001} {\bibfield  {journal} {\bibinfo  {journal} {Physical Review Applied}\ }\textbf {\bibinfo {volume} {5}},\ \bibinfo {pages} {011001} (\bibinfo {year} {2016})}\BibitemShut {NoStop}%
\bibitem [{\citenamefont {Boutin}\ \emph {et~al.}(2017)\citenamefont {Boutin}, \citenamefont {Andersen}, \citenamefont {Venkatraman}, \citenamefont {Ferris},\ and\ \citenamefont {Blais}}]{boutinResonatorResetCircuit2017}%
  \BibitemOpen
  \bibfield  {author} {\bibinfo {author} {\bibfnamefont {S.}~\bibnamefont {Boutin}}, \bibinfo {author} {\bibfnamefont {C.~K.}\ \bibnamefont {Andersen}}, \bibinfo {author} {\bibfnamefont {J.}~\bibnamefont {Venkatraman}}, \bibinfo {author} {\bibfnamefont {A.~J.}\ \bibnamefont {Ferris}},\ and\ \bibinfo {author} {\bibfnamefont {A.}~\bibnamefont {Blais}},\ }\bibfield  {title} {\bibinfo {title} {Resonator reset in circuit {{QED}} by optimal control for large open quantum systems},\ }\href {https://doi.org/10.1103/PhysRevA.96.042315} {\bibfield  {journal} {\bibinfo  {journal} {Physical Review A}\ }\textbf {\bibinfo {volume} {96}},\ \bibinfo {pages} {042315} (\bibinfo {year} {2017})}\BibitemShut {NoStop}%
\bibitem [{\citenamefont {Lled{\'o}}\ \emph {et~al.}(2023)\citenamefont {Lled{\'o}}, \citenamefont {Dassonneville}, \citenamefont {Moulinas}, \citenamefont {Cohen}, \citenamefont {Shillito}, \citenamefont {Bienfait}, \citenamefont {Huard},\ and\ \citenamefont {Blais}}]{lledoCloakingQubitCavity2023}%
  \BibitemOpen
  \bibfield  {author} {\bibinfo {author} {\bibfnamefont {C.}~\bibnamefont {Lled{\'o}}}, \bibinfo {author} {\bibfnamefont {R.}~\bibnamefont {Dassonneville}}, \bibinfo {author} {\bibfnamefont {A.}~\bibnamefont {Moulinas}}, \bibinfo {author} {\bibfnamefont {J.}~\bibnamefont {Cohen}}, \bibinfo {author} {\bibfnamefont {R.}~\bibnamefont {Shillito}}, \bibinfo {author} {\bibfnamefont {A.}~\bibnamefont {Bienfait}}, \bibinfo {author} {\bibfnamefont {B.}~\bibnamefont {Huard}},\ and\ \bibinfo {author} {\bibfnamefont {A.}~\bibnamefont {Blais}},\ }\bibfield  {title} {\bibinfo {title} {Cloaking a qubit in a cavity},\ }\href {https://doi.org/10.1038/s41467-023-42060-5} {\bibfield  {journal} {\bibinfo  {journal} {Nature Communications}\ }\textbf {\bibinfo {volume} {14}},\ \bibinfo {pages} {6313} (\bibinfo {year} {2023})}\BibitemShut {NoStop}%
\bibitem [{\citenamefont {Brillouin}\ and\ \citenamefont {Hellwarth}(1956)}]{brillouinScienceInformationTheory1956}%
  \BibitemOpen
  \bibfield  {author} {\bibinfo {author} {\bibfnamefont {L.}~\bibnamefont {Brillouin}}\ and\ \bibinfo {author} {\bibfnamefont {R.~W.}\ \bibnamefont {Hellwarth}},\ }\bibfield  {title} {\bibinfo {title} {Science and {{Information Theory}}},\ }\href {https://doi.org/10.1063/1.3059856} {\bibfield  {journal} {\bibinfo  {journal} {Physics Today}\ }\textbf {\bibinfo {volume} {9}},\ \bibinfo {pages} {39} (\bibinfo {year} {1956})}\BibitemShut {NoStop}%
\bibitem [{\citenamefont {Groenewold}(1971)}]{groenewoldProblemInformationGain1971}%
  \BibitemOpen
  \bibfield  {author} {\bibinfo {author} {\bibfnamefont {H.~J.}\ \bibnamefont {Groenewold}},\ }\bibfield  {title} {\bibinfo {title} {A problem of information gain by quantal measurements},\ }\href {https://doi.org/10.1007/BF00815357} {\bibfield  {journal} {\bibinfo  {journal} {International Journal of Theoretical Physics}\ }\textbf {\bibinfo {volume} {4}},\ \bibinfo {pages} {327} (\bibinfo {year} {1971})}\BibitemShut {NoStop}%
\bibitem [{\citenamefont {Camati}\ \emph {et~al.}(2016)\citenamefont {Camati}, \citenamefont {Peterson}, \citenamefont {Batalh{\~a}o}, \citenamefont {Micadei}, \citenamefont {Souza}, \citenamefont {Sarthour}, \citenamefont {Oliveira},\ and\ \citenamefont {Serra}}]{camatiExperimentalRectificationEntropy2016}%
  \BibitemOpen
  \bibfield  {author} {\bibinfo {author} {\bibfnamefont {P.~A.}\ \bibnamefont {Camati}}, \bibinfo {author} {\bibfnamefont {J.~P.~S.}\ \bibnamefont {Peterson}}, \bibinfo {author} {\bibfnamefont {T.~B.}\ \bibnamefont {Batalh{\~a}o}}, \bibinfo {author} {\bibfnamefont {K.}~\bibnamefont {Micadei}}, \bibinfo {author} {\bibfnamefont {A.~M.}\ \bibnamefont {Souza}}, \bibinfo {author} {\bibfnamefont {R.~S.}\ \bibnamefont {Sarthour}}, \bibinfo {author} {\bibfnamefont {I.~S.}\ \bibnamefont {Oliveira}},\ and\ \bibinfo {author} {\bibfnamefont {R.~M.}\ \bibnamefont {Serra}},\ }\bibfield  {title} {\bibinfo {title} {Experimental {{Rectification}} of {{Entropy Production}} by {{Maxwell}}'s {{Demon}} in a {{Quantum System}}},\ }\href {https://doi.org/10.1103/PhysRevLett.117.240502} {\bibfield  {journal} {\bibinfo  {journal} {Physical Review Letters}\ }\textbf {\bibinfo {volume} {117}},\ \bibinfo {pages} {240502} (\bibinfo {year} {2016})}\BibitemShut {NoStop}%
\bibitem [{\citenamefont {Pekola}\ \emph {et~al.}(2016)\citenamefont {Pekola}, \citenamefont {Golubev},\ and\ \citenamefont {Averin}}]{pekolaMaxwellDemonBased2016}%
  \BibitemOpen
  \bibfield  {author} {\bibinfo {author} {\bibfnamefont {J.~P.}\ \bibnamefont {Pekola}}, \bibinfo {author} {\bibfnamefont {D.~S.}\ \bibnamefont {Golubev}},\ and\ \bibinfo {author} {\bibfnamefont {D.~V.}\ \bibnamefont {Averin}},\ }\bibfield  {title} {\bibinfo {title} {Maxwell's demon based on a single qubit},\ }\href {https://doi.org/10.1103/PhysRevB.93.024501} {\bibfield  {journal} {\bibinfo  {journal} {Physical Review B}\ }\textbf {\bibinfo {volume} {93}},\ \bibinfo {pages} {024501} (\bibinfo {year} {2016})}\BibitemShut {NoStop}%
\bibitem [{\citenamefont {Naghiloo}\ \emph {et~al.}(2018)\citenamefont {Naghiloo}, \citenamefont {Alonso}, \citenamefont {Romito}, \citenamefont {Lutz},\ and\ \citenamefont {Murch}}]{naghilooInformationGainLoss2018}%
  \BibitemOpen
  \bibfield  {author} {\bibinfo {author} {\bibfnamefont {M.}~\bibnamefont {Naghiloo}}, \bibinfo {author} {\bibfnamefont {J.~J.}\ \bibnamefont {Alonso}}, \bibinfo {author} {\bibfnamefont {A.}~\bibnamefont {Romito}}, \bibinfo {author} {\bibfnamefont {E.}~\bibnamefont {Lutz}},\ and\ \bibinfo {author} {\bibfnamefont {K.~W.}\ \bibnamefont {Murch}},\ }\bibfield  {title} {\bibinfo {title} {Information {{Gain}} and {{Loss}} for a {{Quantum Maxwell}}'s {{Demon}}},\ }\href {https://doi.org/10.1103/PhysRevLett.121.030604} {\bibfield  {journal} {\bibinfo  {journal} {Physical Review Letters}\ }\textbf {\bibinfo {volume} {121}},\ \bibinfo {pages} {030604} (\bibinfo {year} {2018})}\BibitemShut {NoStop}%
\bibitem [{\citenamefont {Masuyama}\ \emph {et~al.}(2018)\citenamefont {Masuyama}, \citenamefont {Funo}, \citenamefont {Murashita}, \citenamefont {Noguchi}, \citenamefont {Kono}, \citenamefont {Tabuchi}, \citenamefont {Yamazaki}, \citenamefont {Ueda},\ and\ \citenamefont {Nakamura}}]{masuyamaInformationtoworkConversionMaxwell2018}%
  \BibitemOpen
  \bibfield  {author} {\bibinfo {author} {\bibfnamefont {Y.}~\bibnamefont {Masuyama}}, \bibinfo {author} {\bibfnamefont {K.}~\bibnamefont {Funo}}, \bibinfo {author} {\bibfnamefont {Y.}~\bibnamefont {Murashita}}, \bibinfo {author} {\bibfnamefont {A.}~\bibnamefont {Noguchi}}, \bibinfo {author} {\bibfnamefont {S.}~\bibnamefont {Kono}}, \bibinfo {author} {\bibfnamefont {Y.}~\bibnamefont {Tabuchi}}, \bibinfo {author} {\bibfnamefont {R.}~\bibnamefont {Yamazaki}}, \bibinfo {author} {\bibfnamefont {M.}~\bibnamefont {Ueda}},\ and\ \bibinfo {author} {\bibfnamefont {Y.}~\bibnamefont {Nakamura}},\ }\bibfield  {title} {\bibinfo {title} {Information-to-work conversion by {{Maxwell}}'s demon in a superconducting circuit quantum electrodynamical system},\ }\href {https://doi.org/10.1038/s41467-018-03686-y} {\bibfield  {journal} {\bibinfo  {journal} {Nature Communications}\ }\textbf {\bibinfo {volume} {9}},\ \bibinfo {pages} {1291} (\bibinfo {year} {2018})}\BibitemShut {NoStop}%
\bibitem [{\citenamefont {Kumar}\ \emph {et~al.}(2018)\citenamefont {Kumar}, \citenamefont {Wu}, \citenamefont {Giraldo},\ and\ \citenamefont {Weiss}}]{kumarSortingUltracoldAtoms2018}%
  \BibitemOpen
  \bibfield  {author} {\bibinfo {author} {\bibfnamefont {A.}~\bibnamefont {Kumar}}, \bibinfo {author} {\bibfnamefont {T.-Y.}\ \bibnamefont {Wu}}, \bibinfo {author} {\bibfnamefont {F.}~\bibnamefont {Giraldo}},\ and\ \bibinfo {author} {\bibfnamefont {D.~S.}\ \bibnamefont {Weiss}},\ }\bibfield  {title} {\bibinfo {title} {Sorting ultracold atoms in a three-dimensional optical lattice in a realization of {{Maxwell}}'s demon},\ }\href {https://doi.org/10.1038/s41586-018-0458-7} {\bibfield  {journal} {\bibinfo  {journal} {Nature}\ }\textbf {\bibinfo {volume} {561}},\ \bibinfo {pages} {83} (\bibinfo {year} {2018})}\BibitemShut {NoStop}%
\bibitem [{\citenamefont {Song}\ \emph {et~al.}(2021)\citenamefont {Song}, \citenamefont {Naghiloo},\ and\ \citenamefont {Murch}}]{songQuantumProcessInference2021}%
  \BibitemOpen
  \bibfield  {author} {\bibinfo {author} {\bibfnamefont {X.}~\bibnamefont {Song}}, \bibinfo {author} {\bibfnamefont {M.}~\bibnamefont {Naghiloo}},\ and\ \bibinfo {author} {\bibfnamefont {K.}~\bibnamefont {Murch}},\ }\bibfield  {title} {\bibinfo {title} {Quantum process inference for a single-qubit {{Maxwell}} demon},\ }\href {https://doi.org/10.1103/PhysRevA.104.022211} {\bibfield  {journal} {\bibinfo  {journal} {Physical Review A}\ }\textbf {\bibinfo {volume} {104}},\ \bibinfo {pages} {022211} (\bibinfo {year} {2021})}\BibitemShut {NoStop}%
\bibitem [{\citenamefont {Kapit}(2017)}]{kapitUpsideNoiseEngineered2017a}%
  \BibitemOpen
  \bibfield  {author} {\bibinfo {author} {\bibfnamefont {E.}~\bibnamefont {Kapit}},\ }\bibfield  {title} {\bibinfo {title} {The upside of noise: Engineered dissipation as a resource in superconducting circuits},\ }\href {https://doi.org/10.1088/2058-9565/aa7e5d} {\bibfield  {journal} {\bibinfo  {journal} {Quantum Science and Technology}\ }\textbf {\bibinfo {volume} {2}},\ \bibinfo {pages} {033002} (\bibinfo {year} {2017})}\BibitemShut {NoStop}%
\bibitem [{\citenamefont {Harrington}\ \emph {et~al.}(2022)\citenamefont {Harrington}, \citenamefont {Mueller},\ and\ \citenamefont {Murch}}]{harringtonEngineeredDissipationQuantum2022}%
  \BibitemOpen
  \bibfield  {author} {\bibinfo {author} {\bibfnamefont {P.~M.}\ \bibnamefont {Harrington}}, \bibinfo {author} {\bibfnamefont {E.~J.}\ \bibnamefont {Mueller}},\ and\ \bibinfo {author} {\bibfnamefont {K.~W.}\ \bibnamefont {Murch}},\ }\bibfield  {title} {\bibinfo {title} {Engineered dissipation for quantum information science},\ }\href {https://doi.org/10.1038/s42254-022-00494-8} {\bibfield  {journal} {\bibinfo  {journal} {Nature Reviews Physics}\ }\textbf {\bibinfo {volume} {4}},\ \bibinfo {pages} {660} (\bibinfo {year} {2022})}\BibitemShut {NoStop}%
\bibitem [{\citenamefont {Valenzuela}\ \emph {et~al.}(2006)\citenamefont {Valenzuela}, \citenamefont {Oliver}, \citenamefont {Berns}, \citenamefont {Berggren}, \citenamefont {Levitov},\ and\ \citenamefont {Orlando}}]{valenzuelaMicrowaveInducedCoolingSuperconducting2006}%
  \BibitemOpen
  \bibfield  {author} {\bibinfo {author} {\bibfnamefont {S.~O.}\ \bibnamefont {Valenzuela}}, \bibinfo {author} {\bibfnamefont {W.~D.}\ \bibnamefont {Oliver}}, \bibinfo {author} {\bibfnamefont {D.~M.}\ \bibnamefont {Berns}}, \bibinfo {author} {\bibfnamefont {K.~K.}\ \bibnamefont {Berggren}}, \bibinfo {author} {\bibfnamefont {L.~S.}\ \bibnamefont {Levitov}},\ and\ \bibinfo {author} {\bibfnamefont {T.~P.}\ \bibnamefont {Orlando}},\ }\bibfield  {title} {\bibinfo {title} {Microwave-{{Induced Cooling}} of a {{Superconducting Qubit}}},\ }\href {https://doi.org/10.1126/science.1134008} {\bibfield  {journal} {\bibinfo  {journal} {Science}\ }\textbf {\bibinfo {volume} {314}},\ \bibinfo {pages} {1589} (\bibinfo {year} {2006})}\BibitemShut {NoStop}%
\bibitem [{\citenamefont {Geerlings}\ \emph {et~al.}(2013)\citenamefont {Geerlings}, \citenamefont {Leghtas}, \citenamefont {Pop}, \citenamefont {Shankar}, \citenamefont {Frunzio}, \citenamefont {Schoelkopf}, \citenamefont {Mirrahimi},\ and\ \citenamefont {Devoret}}]{geerlingsDemonstratingDrivenReset2013}%
  \BibitemOpen
  \bibfield  {author} {\bibinfo {author} {\bibfnamefont {K.}~\bibnamefont {Geerlings}}, \bibinfo {author} {\bibfnamefont {Z.}~\bibnamefont {Leghtas}}, \bibinfo {author} {\bibfnamefont {I.~M.}\ \bibnamefont {Pop}}, \bibinfo {author} {\bibfnamefont {S.}~\bibnamefont {Shankar}}, \bibinfo {author} {\bibfnamefont {L.}~\bibnamefont {Frunzio}}, \bibinfo {author} {\bibfnamefont {R.~J.}\ \bibnamefont {Schoelkopf}}, \bibinfo {author} {\bibfnamefont {M.}~\bibnamefont {Mirrahimi}},\ and\ \bibinfo {author} {\bibfnamefont {M.~H.}\ \bibnamefont {Devoret}},\ }\bibfield  {title} {\bibinfo {title} {Demonstrating a {{Driven Reset Protocol}} for a {{Superconducting Qubit}}},\ }\href {https://doi.org/10.1103/PhysRevLett.110.120501} {\bibfield  {journal} {\bibinfo  {journal} {Physical Review Letters}\ }\textbf {\bibinfo {volume} {110}},\ \bibinfo {pages} {120501} (\bibinfo {year} {2013})}\BibitemShut {NoStop}%
\bibitem [{\citenamefont {Magnard}\ \emph {et~al.}(2018)\citenamefont {Magnard}, \citenamefont {Kurpiers}, \citenamefont {Royer}, \citenamefont {Walter}, \citenamefont {Besse}, \citenamefont {Gasparinetti}, \citenamefont {Pechal}, \citenamefont {Heinsoo}, \citenamefont {Storz}, \citenamefont {Blais},\ and\ \citenamefont {Wallraff}}]{magnardFastUnconditionalAllMicrowave2018}%
  \BibitemOpen
  \bibfield  {author} {\bibinfo {author} {\bibfnamefont {P.}~\bibnamefont {Magnard}}, \bibinfo {author} {\bibfnamefont {P.}~\bibnamefont {Kurpiers}}, \bibinfo {author} {\bibfnamefont {B.}~\bibnamefont {Royer}}, \bibinfo {author} {\bibfnamefont {T.}~\bibnamefont {Walter}}, \bibinfo {author} {\bibfnamefont {J.-C.}\ \bibnamefont {Besse}}, \bibinfo {author} {\bibfnamefont {S.}~\bibnamefont {Gasparinetti}}, \bibinfo {author} {\bibfnamefont {M.}~\bibnamefont {Pechal}}, \bibinfo {author} {\bibfnamefont {J.}~\bibnamefont {Heinsoo}}, \bibinfo {author} {\bibfnamefont {S.}~\bibnamefont {Storz}}, \bibinfo {author} {\bibfnamefont {A.}~\bibnamefont {Blais}},\ and\ \bibinfo {author} {\bibfnamefont {A.}~\bibnamefont {Wallraff}},\ }\bibfield  {title} {\bibinfo {title} {Fast and {{Unconditional All-Microwave Reset}} of a {{Superconducting Qubit}}},\ }\href {https://doi.org/10.1103/PhysRevLett.121.060502} {\bibfield  {journal} {\bibinfo  {journal} {Physical Review Letters}\ }\textbf {\bibinfo {volume} {121}},\ \bibinfo {pages}
  {060502} (\bibinfo {year} {2018})}\BibitemShut {NoStop}%
\bibitem [{\citenamefont {Egger}\ \emph {et~al.}(2018)\citenamefont {Egger}, \citenamefont {Werninghaus}, \citenamefont {Ganzhorn}, \citenamefont {Salis}, \citenamefont {Fuhrer}, \citenamefont {M{\"u}ller},\ and\ \citenamefont {Filipp}}]{eggerPulsedResetProtocol2018}%
  \BibitemOpen
  \bibfield  {author} {\bibinfo {author} {\bibfnamefont {D.}~\bibnamefont {Egger}}, \bibinfo {author} {\bibfnamefont {M.}~\bibnamefont {Werninghaus}}, \bibinfo {author} {\bibfnamefont {M.}~\bibnamefont {Ganzhorn}}, \bibinfo {author} {\bibfnamefont {G.}~\bibnamefont {Salis}}, \bibinfo {author} {\bibfnamefont {A.}~\bibnamefont {Fuhrer}}, \bibinfo {author} {\bibfnamefont {P.}~\bibnamefont {M{\"u}ller}},\ and\ \bibinfo {author} {\bibfnamefont {S.}~\bibnamefont {Filipp}},\ }\bibfield  {title} {\bibinfo {title} {Pulsed {{Reset Protocol}} for {{Fixed-Frequency Superconducting Qubits}}},\ }\href {https://doi.org/10.1103/PhysRevApplied.10.044030} {\bibfield  {journal} {\bibinfo  {journal} {Physical Review Applied}\ }\textbf {\bibinfo {volume} {10}},\ \bibinfo {pages} {044030} (\bibinfo {year} {2018})}\BibitemShut {NoStop}%
\bibitem [{\citenamefont {Zhou}\ \emph {et~al.}(2021)\citenamefont {Zhou}, \citenamefont {Zhang}, \citenamefont {Yin}, \citenamefont {Huai}, \citenamefont {Gu}, \citenamefont {Xu}, \citenamefont {Allcock}, \citenamefont {Liu}, \citenamefont {Xi}, \citenamefont {Yu}, \citenamefont {Zhang}, \citenamefont {Zhang}, \citenamefont {Li}, \citenamefont {Song}, \citenamefont {Wang}, \citenamefont {Zheng}, \citenamefont {An}, \citenamefont {Zheng},\ and\ \citenamefont {Zhang}}]{zhouRapidUnconditionalParametric2021a}%
  \BibitemOpen
  \bibfield  {author} {\bibinfo {author} {\bibfnamefont {Y.}~\bibnamefont {Zhou}}, \bibinfo {author} {\bibfnamefont {Z.}~\bibnamefont {Zhang}}, \bibinfo {author} {\bibfnamefont {Z.}~\bibnamefont {Yin}}, \bibinfo {author} {\bibfnamefont {S.}~\bibnamefont {Huai}}, \bibinfo {author} {\bibfnamefont {X.}~\bibnamefont {Gu}}, \bibinfo {author} {\bibfnamefont {X.}~\bibnamefont {Xu}}, \bibinfo {author} {\bibfnamefont {J.}~\bibnamefont {Allcock}}, \bibinfo {author} {\bibfnamefont {F.}~\bibnamefont {Liu}}, \bibinfo {author} {\bibfnamefont {G.}~\bibnamefont {Xi}}, \bibinfo {author} {\bibfnamefont {Q.}~\bibnamefont {Yu}}, \bibinfo {author} {\bibfnamefont {H.}~\bibnamefont {Zhang}}, \bibinfo {author} {\bibfnamefont {M.}~\bibnamefont {Zhang}}, \bibinfo {author} {\bibfnamefont {H.}~\bibnamefont {Li}}, \bibinfo {author} {\bibfnamefont {X.}~\bibnamefont {Song}}, \bibinfo {author} {\bibfnamefont {Z.}~\bibnamefont {Wang}}, \bibinfo {author} {\bibfnamefont {D.}~\bibnamefont {Zheng}}, \bibinfo {author} {\bibfnamefont
  {S.}~\bibnamefont {An}}, \bibinfo {author} {\bibfnamefont {Y.}~\bibnamefont {Zheng}},\ and\ \bibinfo {author} {\bibfnamefont {S.}~\bibnamefont {Zhang}},\ }\bibfield  {title} {\bibinfo {title} {Rapid and unconditional parametric reset protocol for tunable superconducting qubits},\ }\href {https://doi.org/10.1038/s41467-021-26205-y} {\bibfield  {journal} {\bibinfo  {journal} {Nature Communications}\ }\textbf {\bibinfo {volume} {12}},\ \bibinfo {pages} {5924} (\bibinfo {year} {2021})}\BibitemShut {NoStop}%
\bibitem [{\citenamefont {Lacroix}\ \emph {et~al.}(2023)\citenamefont {Lacroix}, \citenamefont {Hofele}, \citenamefont {Remm}, \citenamefont {{Benhayoune-Khadraoui}}, \citenamefont {McDonald}, \citenamefont {Shillito}, \citenamefont {Lazar}, \citenamefont {Hellings}, \citenamefont {Swiadek}, \citenamefont {{Colao-Zanuz}}, \citenamefont {Flasby}, \citenamefont {Panah}, \citenamefont {Kerschbaum}, \citenamefont {Norris}, \citenamefont {Blais}, \citenamefont {Wallraff},\ and\ \citenamefont {Krinner}}]{lacroixFastFluxActivatedLeakage2023}%
  \BibitemOpen
  \bibfield  {author} {\bibinfo {author} {\bibfnamefont {N.}~\bibnamefont {Lacroix}}, \bibinfo {author} {\bibfnamefont {L.}~\bibnamefont {Hofele}}, \bibinfo {author} {\bibfnamefont {A.}~\bibnamefont {Remm}}, \bibinfo {author} {\bibfnamefont {O.}~\bibnamefont {{Benhayoune-Khadraoui}}}, \bibinfo {author} {\bibfnamefont {A.}~\bibnamefont {McDonald}}, \bibinfo {author} {\bibfnamefont {R.}~\bibnamefont {Shillito}}, \bibinfo {author} {\bibfnamefont {S.}~\bibnamefont {Lazar}}, \bibinfo {author} {\bibfnamefont {C.}~\bibnamefont {Hellings}}, \bibinfo {author} {\bibfnamefont {F.}~\bibnamefont {Swiadek}}, \bibinfo {author} {\bibfnamefont {D.}~\bibnamefont {{Colao-Zanuz}}}, \bibinfo {author} {\bibfnamefont {A.}~\bibnamefont {Flasby}}, \bibinfo {author} {\bibfnamefont {M.~B.}\ \bibnamefont {Panah}}, \bibinfo {author} {\bibfnamefont {M.}~\bibnamefont {Kerschbaum}}, \bibinfo {author} {\bibfnamefont {G.~J.}\ \bibnamefont {Norris}}, \bibinfo {author} {\bibfnamefont {A.}~\bibnamefont {Blais}}, \bibinfo {author} {\bibfnamefont
  {A.}~\bibnamefont {Wallraff}},\ and\ \bibinfo {author} {\bibfnamefont {S.}~\bibnamefont {Krinner}},\ }\href {https://doi.org/10.48550/arXiv.2309.07060} {\bibinfo {title} {Fast {{Flux-Activated Leakage Reduction}} for {{Superconducting Quantum Circuits}}}} (\bibinfo {year} {2023}),\ \Eprint {https://arxiv.org/abs/2309.07060} {arxiv:2309.07060 [quant-ph]} \BibitemShut {NoStop}%
\bibitem [{\citenamefont {Murch}\ \emph {et~al.}(2012)\citenamefont {Murch}, \citenamefont {Vool}, \citenamefont {Zhou}, \citenamefont {Weber}, \citenamefont {Girvin},\ and\ \citenamefont {Siddiqi}}]{murchCavityAssistedQuantumBath2012}%
  \BibitemOpen
  \bibfield  {author} {\bibinfo {author} {\bibfnamefont {K.~W.}\ \bibnamefont {Murch}}, \bibinfo {author} {\bibfnamefont {U.}~\bibnamefont {Vool}}, \bibinfo {author} {\bibfnamefont {D.}~\bibnamefont {Zhou}}, \bibinfo {author} {\bibfnamefont {S.~J.}\ \bibnamefont {Weber}}, \bibinfo {author} {\bibfnamefont {S.~M.}\ \bibnamefont {Girvin}},\ and\ \bibinfo {author} {\bibfnamefont {I.}~\bibnamefont {Siddiqi}},\ }\bibfield  {title} {\bibinfo {title} {Cavity-{{Assisted Quantum Bath Engineering}}},\ }\href {https://doi.org/10.1103/PhysRevLett.109.183602} {\bibfield  {journal} {\bibinfo  {journal} {Physical Review Letters}\ }\textbf {\bibinfo {volume} {109}},\ \bibinfo {pages} {183602} (\bibinfo {year} {2012})}\BibitemShut {NoStop}%
\bibitem [{\citenamefont {Lu}\ \emph {et~al.}(2017)\citenamefont {Lu}, \citenamefont {Chakram}, \citenamefont {Leung}, \citenamefont {Earnest}, \citenamefont {Naik}, \citenamefont {Huang}, \citenamefont {Groszkowski}, \citenamefont {Kapit}, \citenamefont {Koch},\ and\ \citenamefont {Schuster}}]{luUniversalStabilizationParametrically2017a}%
  \BibitemOpen
  \bibfield  {author} {\bibinfo {author} {\bibfnamefont {Y.}~\bibnamefont {Lu}}, \bibinfo {author} {\bibfnamefont {S.}~\bibnamefont {Chakram}}, \bibinfo {author} {\bibfnamefont {N.}~\bibnamefont {Leung}}, \bibinfo {author} {\bibfnamefont {N.}~\bibnamefont {Earnest}}, \bibinfo {author} {\bibfnamefont {R.~K.}\ \bibnamefont {Naik}}, \bibinfo {author} {\bibfnamefont {Z.}~\bibnamefont {Huang}}, \bibinfo {author} {\bibfnamefont {P.}~\bibnamefont {Groszkowski}}, \bibinfo {author} {\bibfnamefont {E.}~\bibnamefont {Kapit}}, \bibinfo {author} {\bibfnamefont {J.}~\bibnamefont {Koch}},\ and\ \bibinfo {author} {\bibfnamefont {D.~I.}\ \bibnamefont {Schuster}},\ }\bibfield  {title} {\bibinfo {title} {Universal {{Stabilization}} of a {{Parametrically Coupled Qubit}}},\ }\href {https://doi.org/10.1103/PhysRevLett.119.150502} {\bibfield  {journal} {\bibinfo  {journal} {Physical Review Letters}\ }\textbf {\bibinfo {volume} {119}},\ \bibinfo {pages} {150502} (\bibinfo {year} {2017})}\BibitemShut {NoStop}%
\bibitem [{\citenamefont {Shankar}\ \emph {et~al.}(2013)\citenamefont {Shankar}, \citenamefont {Hatridge}, \citenamefont {Leghtas}, \citenamefont {Sliwa}, \citenamefont {Narla}, \citenamefont {Vool}, \citenamefont {Girvin}, \citenamefont {Frunzio}, \citenamefont {Mirrahimi},\ and\ \citenamefont {Devoret}}]{shankarAutonomouslyStabilizedEntanglement2013}%
  \BibitemOpen
  \bibfield  {author} {\bibinfo {author} {\bibfnamefont {S.}~\bibnamefont {Shankar}}, \bibinfo {author} {\bibfnamefont {M.}~\bibnamefont {Hatridge}}, \bibinfo {author} {\bibfnamefont {Z.}~\bibnamefont {Leghtas}}, \bibinfo {author} {\bibfnamefont {K.~M.}\ \bibnamefont {Sliwa}}, \bibinfo {author} {\bibfnamefont {A.}~\bibnamefont {Narla}}, \bibinfo {author} {\bibfnamefont {U.}~\bibnamefont {Vool}}, \bibinfo {author} {\bibfnamefont {S.~M.}\ \bibnamefont {Girvin}}, \bibinfo {author} {\bibfnamefont {L.}~\bibnamefont {Frunzio}}, \bibinfo {author} {\bibfnamefont {M.}~\bibnamefont {Mirrahimi}},\ and\ \bibinfo {author} {\bibfnamefont {M.~H.}\ \bibnamefont {Devoret}},\ }\bibfield  {title} {\bibinfo {title} {Autonomously stabilized entanglement between two superconducting quantum bits},\ }\href {https://doi.org/10.1038/nature12802} {\bibfield  {journal} {\bibinfo  {journal} {Nature}\ }\textbf {\bibinfo {volume} {504}},\ \bibinfo {pages} {419} (\bibinfo {year} {2013})}\BibitemShut {NoStop}%
\bibitem [{\citenamefont {{Kimchi-Schwartz}}\ \emph {et~al.}(2016)\citenamefont {{Kimchi-Schwartz}}, \citenamefont {Martin}, \citenamefont {Flurin}, \citenamefont {Aron}, \citenamefont {Kulkarni}, \citenamefont {Tureci},\ and\ \citenamefont {Siddiqi}}]{kimchi-schwartzStabilizingEntanglementSymmetrySelective2016a}%
  \BibitemOpen
  \bibfield  {author} {\bibinfo {author} {\bibfnamefont {M.~E.}\ \bibnamefont {{Kimchi-Schwartz}}}, \bibinfo {author} {\bibfnamefont {L.}~\bibnamefont {Martin}}, \bibinfo {author} {\bibfnamefont {E.}~\bibnamefont {Flurin}}, \bibinfo {author} {\bibfnamefont {C.}~\bibnamefont {Aron}}, \bibinfo {author} {\bibfnamefont {M.}~\bibnamefont {Kulkarni}}, \bibinfo {author} {\bibfnamefont {H.~E.}\ \bibnamefont {Tureci}},\ and\ \bibinfo {author} {\bibfnamefont {I.}~\bibnamefont {Siddiqi}},\ }\bibfield  {title} {\bibinfo {title} {Stabilizing {{Entanglement}} via {{Symmetry-Selective Bath Engineering}} in {{Superconducting Qubits}}},\ }\href {https://doi.org/10.1103/PhysRevLett.116.240503} {\bibfield  {journal} {\bibinfo  {journal} {Physical Review Letters}\ }\textbf {\bibinfo {volume} {116}},\ \bibinfo {pages} {240503} (\bibinfo {year} {2016})}\BibitemShut {NoStop}%
\bibitem [{\citenamefont {Holland}\ \emph {et~al.}(2015)\citenamefont {Holland}, \citenamefont {Vlastakis}, \citenamefont {Heeres}, \citenamefont {Reagor}, \citenamefont {Vool}, \citenamefont {Leghtas}, \citenamefont {Frunzio}, \citenamefont {Kirchmair}, \citenamefont {Devoret}, \citenamefont {Mirrahimi},\ and\ \citenamefont {Schoelkopf}}]{hollandSinglePhotonResolvedCrossKerrInteraction2015}%
  \BibitemOpen
  \bibfield  {author} {\bibinfo {author} {\bibfnamefont {E.~T.}\ \bibnamefont {Holland}}, \bibinfo {author} {\bibfnamefont {B.}~\bibnamefont {Vlastakis}}, \bibinfo {author} {\bibfnamefont {R.~W.}\ \bibnamefont {Heeres}}, \bibinfo {author} {\bibfnamefont {M.~J.}\ \bibnamefont {Reagor}}, \bibinfo {author} {\bibfnamefont {U.}~\bibnamefont {Vool}}, \bibinfo {author} {\bibfnamefont {Z.}~\bibnamefont {Leghtas}}, \bibinfo {author} {\bibfnamefont {L.}~\bibnamefont {Frunzio}}, \bibinfo {author} {\bibfnamefont {G.}~\bibnamefont {Kirchmair}}, \bibinfo {author} {\bibfnamefont {M.~H.}\ \bibnamefont {Devoret}}, \bibinfo {author} {\bibfnamefont {M.}~\bibnamefont {Mirrahimi}},\ and\ \bibinfo {author} {\bibfnamefont {R.~J.}\ \bibnamefont {Schoelkopf}},\ }\bibfield  {title} {\bibinfo {title} {Single-{{Photon-Resolved Cross-Kerr Interaction}} for {{Autonomous Stabilization}} of {{Photon-Number States}}},\ }\href {https://doi.org/10.1103/PhysRevLett.115.180501} {\bibfield  {journal} {\bibinfo  {journal} {Physical Review
  Letters}\ }\textbf {\bibinfo {volume} {115}},\ \bibinfo {pages} {180501} (\bibinfo {year} {2015})}\BibitemShut {NoStop}%
\bibitem [{\citenamefont {Leghtas}\ \emph {et~al.}(2015)\citenamefont {Leghtas}, \citenamefont {Touzard}, \citenamefont {Pop}, \citenamefont {Kou}, \citenamefont {Vlastakis}, \citenamefont {Petrenko}, \citenamefont {Sliwa}, \citenamefont {Narla}, \citenamefont {Shankar}, \citenamefont {Hatridge}, \citenamefont {Reagor}, \citenamefont {Frunzio}, \citenamefont {Schoelkopf}, \citenamefont {Mirrahimi},\ and\ \citenamefont {Devoret}}]{leghtasConfiningStateLight2015}%
  \BibitemOpen
  \bibfield  {author} {\bibinfo {author} {\bibfnamefont {Z.}~\bibnamefont {Leghtas}}, \bibinfo {author} {\bibfnamefont {S.}~\bibnamefont {Touzard}}, \bibinfo {author} {\bibfnamefont {I.~M.}\ \bibnamefont {Pop}}, \bibinfo {author} {\bibfnamefont {A.}~\bibnamefont {Kou}}, \bibinfo {author} {\bibfnamefont {B.}~\bibnamefont {Vlastakis}}, \bibinfo {author} {\bibfnamefont {A.}~\bibnamefont {Petrenko}}, \bibinfo {author} {\bibfnamefont {K.~M.}\ \bibnamefont {Sliwa}}, \bibinfo {author} {\bibfnamefont {A.}~\bibnamefont {Narla}}, \bibinfo {author} {\bibfnamefont {S.}~\bibnamefont {Shankar}}, \bibinfo {author} {\bibfnamefont {M.~J.}\ \bibnamefont {Hatridge}}, \bibinfo {author} {\bibfnamefont {M.}~\bibnamefont {Reagor}}, \bibinfo {author} {\bibfnamefont {L.}~\bibnamefont {Frunzio}}, \bibinfo {author} {\bibfnamefont {R.~J.}\ \bibnamefont {Schoelkopf}}, \bibinfo {author} {\bibfnamefont {M.}~\bibnamefont {Mirrahimi}},\ and\ \bibinfo {author} {\bibfnamefont {M.~H.}\ \bibnamefont {Devoret}},\ }\bibfield  {title} {\bibinfo
  {title} {Confining the state of light to a quantum manifold by engineered two-photon loss},\ }\href {https://doi.org/10.1126/science.aaa2085} {\bibfield  {journal} {\bibinfo  {journal} {Science}\ }\textbf {\bibinfo {volume} {347}},\ \bibinfo {pages} {853} (\bibinfo {year} {2015})}\BibitemShut {NoStop}%
\bibitem [{\citenamefont {Wong}\ \emph {et~al.}(2019)\citenamefont {Wong}, \citenamefont {Wilen}, \citenamefont {McDermott},\ and\ \citenamefont {Vavilov}}]{wongTunableQuantumDissipator2019}%
  \BibitemOpen
  \bibfield  {author} {\bibinfo {author} {\bibfnamefont {C.~H.}\ \bibnamefont {Wong}}, \bibinfo {author} {\bibfnamefont {C.}~\bibnamefont {Wilen}}, \bibinfo {author} {\bibfnamefont {R.}~\bibnamefont {McDermott}},\ and\ \bibinfo {author} {\bibfnamefont {M.~G.}\ \bibnamefont {Vavilov}},\ }\bibfield  {title} {\bibinfo {title} {A tunable quantum dissipator for active resonator reset in circuit {{QED}}},\ }\href {https://doi.org/10.1088/2058-9565/aaf6d3} {\bibfield  {journal} {\bibinfo  {journal} {Quantum Science and Technology}\ }\textbf {\bibinfo {volume} {4}},\ \bibinfo {pages} {025001} (\bibinfo {year} {2019})}\BibitemShut {NoStop}%
\bibitem [{\citenamefont {Lu}\ \emph {et~al.}(2023)\citenamefont {Lu}, \citenamefont {Maiti}, \citenamefont {Garmon}, \citenamefont {Ganjam}, \citenamefont {Zhang}, \citenamefont {Claes}, \citenamefont {Frunzio}, \citenamefont {Girvin},\ and\ \citenamefont {Schoelkopf}}]{luHighfidelityParametricBeamsplitting2023}%
  \BibitemOpen
  \bibfield  {author} {\bibinfo {author} {\bibfnamefont {Y.}~\bibnamefont {Lu}}, \bibinfo {author} {\bibfnamefont {A.}~\bibnamefont {Maiti}}, \bibinfo {author} {\bibfnamefont {J.~W.~O.}\ \bibnamefont {Garmon}}, \bibinfo {author} {\bibfnamefont {S.}~\bibnamefont {Ganjam}}, \bibinfo {author} {\bibfnamefont {Y.}~\bibnamefont {Zhang}}, \bibinfo {author} {\bibfnamefont {J.}~\bibnamefont {Claes}}, \bibinfo {author} {\bibfnamefont {L.}~\bibnamefont {Frunzio}}, \bibinfo {author} {\bibfnamefont {S.~M.}\ \bibnamefont {Girvin}},\ and\ \bibinfo {author} {\bibfnamefont {R.~J.}\ \bibnamefont {Schoelkopf}},\ }\bibfield  {title} {\bibinfo {title} {High-fidelity parametric beamsplitting with a parity-protected converter},\ }\href {https://doi.org/10.1038/s41467-023-41104-0} {\bibfield  {journal} {\bibinfo  {journal} {Nature Communications}\ }\textbf {\bibinfo {volume} {14}},\ \bibinfo {pages} {5767} (\bibinfo {year} {2023})}\BibitemShut {NoStop}%
\bibitem [{Note1()}]{Note1}%
  \BibitemOpen
  \bibinfo {note} {This is the regime in which we operate our device, based on the measured parameters. Note that if we were not in this regime, during modulation the dissipator frequency would come near resonance with the cavity, and the two would undergo a direct exchange operation. This may produce the desired loss, but such frequency collisions are exactly what we sought to avoid with parametric drives, as there may be other modes near the cavity that we do not wish to make lossy.}\BibitemShut {Stop}%
\bibitem [{\citenamefont {Okamoto}\ \emph {et~al.}(2013)\citenamefont {Okamoto}, \citenamefont {Gourgout}, \citenamefont {Chang}, \citenamefont {Onomitsu}, \citenamefont {Mahboob}, \citenamefont {Chang},\ and\ \citenamefont {Yamaguchi}}]{okamotoCoherentPhononManipulation2013}%
  \BibitemOpen
  \bibfield  {author} {\bibinfo {author} {\bibfnamefont {H.}~\bibnamefont {Okamoto}}, \bibinfo {author} {\bibfnamefont {A.}~\bibnamefont {Gourgout}}, \bibinfo {author} {\bibfnamefont {C.-Y.}\ \bibnamefont {Chang}}, \bibinfo {author} {\bibfnamefont {K.}~\bibnamefont {Onomitsu}}, \bibinfo {author} {\bibfnamefont {I.}~\bibnamefont {Mahboob}}, \bibinfo {author} {\bibfnamefont {E.~Y.}\ \bibnamefont {Chang}},\ and\ \bibinfo {author} {\bibfnamefont {H.}~\bibnamefont {Yamaguchi}},\ }\bibfield  {title} {\bibinfo {title} {Coherent phonon manipulation in coupled mechanical resonators},\ }\href {https://doi.org/10.1038/nphys2665} {\bibfield  {journal} {\bibinfo  {journal} {Nature Physics}\ }\textbf {\bibinfo {volume} {9}},\ \bibinfo {pages} {480} (\bibinfo {year} {2013})}\BibitemShut {NoStop}%
\bibitem [{\citenamefont {Doucet}\ \emph {et~al.}(2020)\citenamefont {Doucet}, \citenamefont {Reiter}, \citenamefont {Ranzani},\ and\ \citenamefont {Kamal}}]{doucetHighFidelityDissipation2020}%
  \BibitemOpen
  \bibfield  {author} {\bibinfo {author} {\bibfnamefont {E.}~\bibnamefont {Doucet}}, \bibinfo {author} {\bibfnamefont {F.}~\bibnamefont {Reiter}}, \bibinfo {author} {\bibfnamefont {L.}~\bibnamefont {Ranzani}},\ and\ \bibinfo {author} {\bibfnamefont {A.}~\bibnamefont {Kamal}},\ }\bibfield  {title} {\bibinfo {title} {High fidelity dissipation engineering using parametric interactions},\ }\href {https://doi.org/10.1103/PhysRevResearch.2.023370} {\bibfield  {journal} {\bibinfo  {journal} {Physical Review Research}\ }\textbf {\bibinfo {volume} {2}},\ \bibinfo {pages} {023370} (\bibinfo {year} {2020})}\BibitemShut {NoStop}%
\bibitem [{\citenamefont {Zhou}\ \emph {et~al.}(2024)\citenamefont {Zhou}, \citenamefont {Cai}, \citenamefont {Zheng}, \citenamefont {Zhou}, \citenamefont {Wang}, \citenamefont {Xiong},\ and\ \citenamefont {Feng}}]{zhouHighsuppressionratioWideBandwidth2024}%
  \BibitemOpen
  \bibfield  {author} {\bibinfo {author} {\bibfnamefont {Y.}~\bibnamefont {Zhou}}, \bibinfo {author} {\bibfnamefont {X.}~\bibnamefont {Cai}}, \bibinfo {author} {\bibfnamefont {Y.}~\bibnamefont {Zheng}}, \bibinfo {author} {\bibfnamefont {B.}~\bibnamefont {Zhou}}, \bibinfo {author} {\bibfnamefont {Y.}~\bibnamefont {Wang}}, \bibinfo {author} {\bibfnamefont {K.}~\bibnamefont {Xiong}},\ and\ \bibinfo {author} {\bibfnamefont {J.}~\bibnamefont {Feng}},\ }\bibfield  {title} {\bibinfo {title} {High-suppression-ratio and wide bandwidth four-stage {{Purcell}} filter for multiplexed superconducting qubit readout},\ }\href {https://doi.org/10.1063/5.0173539} {\bibfield  {journal} {\bibinfo  {journal} {Journal of Applied Physics}\ }\textbf {\bibinfo {volume} {135}},\ \bibinfo {pages} {024402} (\bibinfo {year} {2024})}\BibitemShut {NoStop}%
\bibitem [{\citenamefont {Gertler}\ \emph {et~al.}(2021)\citenamefont {Gertler}, \citenamefont {Baker}, \citenamefont {Li}, \citenamefont {Shirol}, \citenamefont {Koch},\ and\ \citenamefont {Wang}}]{gertlerProtectingBosonicQubit2021}%
  \BibitemOpen
  \bibfield  {author} {\bibinfo {author} {\bibfnamefont {J.~M.}\ \bibnamefont {Gertler}}, \bibinfo {author} {\bibfnamefont {B.}~\bibnamefont {Baker}}, \bibinfo {author} {\bibfnamefont {J.}~\bibnamefont {Li}}, \bibinfo {author} {\bibfnamefont {S.}~\bibnamefont {Shirol}}, \bibinfo {author} {\bibfnamefont {J.}~\bibnamefont {Koch}},\ and\ \bibinfo {author} {\bibfnamefont {C.}~\bibnamefont {Wang}},\ }\bibfield  {title} {\bibinfo {title} {Protecting a bosonic qubit with autonomous quantum error correction},\ }\href {https://doi.org/10.1038/s41586-021-03257-0} {\bibfield  {journal} {\bibinfo  {journal} {Nature}\ }\textbf {\bibinfo {volume} {590}},\ \bibinfo {pages} {243} (\bibinfo {year} {2021})}\BibitemShut {NoStop}%
\bibitem [{\citenamefont {Kudra}\ \emph {et~al.}(2022)\citenamefont {Kudra}, \citenamefont {Abad}, \citenamefont {Kervinen}, \citenamefont {Eriksson}, \citenamefont {Quijandr{\'i}a}, \citenamefont {Delsing},\ and\ \citenamefont {Gasparinetti}}]{kudraExperimentalRealizationDeterministic2022}%
  \BibitemOpen
  \bibfield  {author} {\bibinfo {author} {\bibfnamefont {M.}~\bibnamefont {Kudra}}, \bibinfo {author} {\bibfnamefont {T.}~\bibnamefont {Abad}}, \bibinfo {author} {\bibfnamefont {M.}~\bibnamefont {Kervinen}}, \bibinfo {author} {\bibfnamefont {A.~M.}\ \bibnamefont {Eriksson}}, \bibinfo {author} {\bibfnamefont {F.}~\bibnamefont {Quijandr{\'i}a}}, \bibinfo {author} {\bibfnamefont {P.}~\bibnamefont {Delsing}},\ and\ \bibinfo {author} {\bibfnamefont {S.}~\bibnamefont {Gasparinetti}},\ }\href {https://doi.org/10.48550/arXiv.2212.12079} {\bibinfo {title} {Experimental realization of deterministic and selective photon addition in a bosonic mode assisted by an ancillary qubit}} (\bibinfo {year} {2022}),\ \Eprint {https://arxiv.org/abs/2212.12079} {arxiv:2212.12079 [quant-ph]} \BibitemShut {NoStop}%
\bibitem [{\citenamefont {{Lachance-Quirion}}\ \emph {et~al.}(2023)\citenamefont {{Lachance-Quirion}}, \citenamefont {Lemonde}, \citenamefont {Simoneau}, \citenamefont {{St-Jean}}, \citenamefont {Lemieux}, \citenamefont {Turcotte}, \citenamefont {Wright}, \citenamefont {Lacroix}, \citenamefont {{Fr{\'e}chette-Viens}}, \citenamefont {Shillito}, \citenamefont {Hopfmueller}, \citenamefont {Tremblay}, \citenamefont {Frattini}, \citenamefont {Lemyre},\ and\ \citenamefont {{St-Jean}}}]{lachance-quirionAutonomousQuantumError2023}%
  \BibitemOpen
  \bibfield  {author} {\bibinfo {author} {\bibfnamefont {D.}~\bibnamefont {{Lachance-Quirion}}}, \bibinfo {author} {\bibfnamefont {M.-A.}\ \bibnamefont {Lemonde}}, \bibinfo {author} {\bibfnamefont {J.~O.}\ \bibnamefont {Simoneau}}, \bibinfo {author} {\bibfnamefont {L.}~\bibnamefont {{St-Jean}}}, \bibinfo {author} {\bibfnamefont {P.}~\bibnamefont {Lemieux}}, \bibinfo {author} {\bibfnamefont {S.}~\bibnamefont {Turcotte}}, \bibinfo {author} {\bibfnamefont {W.}~\bibnamefont {Wright}}, \bibinfo {author} {\bibfnamefont {A.}~\bibnamefont {Lacroix}}, \bibinfo {author} {\bibfnamefont {J.}~\bibnamefont {{Fr{\'e}chette-Viens}}}, \bibinfo {author} {\bibfnamefont {R.}~\bibnamefont {Shillito}}, \bibinfo {author} {\bibfnamefont {F.}~\bibnamefont {Hopfmueller}}, \bibinfo {author} {\bibfnamefont {M.}~\bibnamefont {Tremblay}}, \bibinfo {author} {\bibfnamefont {N.~E.}\ \bibnamefont {Frattini}}, \bibinfo {author} {\bibfnamefont {J.~C.}\ \bibnamefont {Lemyre}},\ and\ \bibinfo {author} {\bibfnamefont {P.}~\bibnamefont
  {{St-Jean}}},\ }\href {https://doi.org/10.48550/arXiv.2310.11400} {\bibinfo {title} {Autonomous quantum error correction of {{Gottesman-Kitaev-Preskill}} states}} (\bibinfo {year} {2023}),\ \Eprint {https://arxiv.org/abs/2310.11400} {arxiv:2310.11400 [cond-mat, physics:quant-ph]} \BibitemShut {NoStop}%
\bibitem [{\citenamefont {Mi}\ \emph {et~al.}(2023)\citenamefont {Mi}, \citenamefont {Michailidis}, \citenamefont {Shabani}, \citenamefont {Miao}, \citenamefont {Klimov}, \citenamefont {Lloyd}, \citenamefont {Rosenberg}, \citenamefont {Acharya}, \citenamefont {Aleiner}, \citenamefont {Andersen}, \citenamefont {Ansmann}, \citenamefont {Arute}, \citenamefont {Arya}, \citenamefont {Asfaw}, \citenamefont {Atalaya}, \citenamefont {Bardin}, \citenamefont {Bengtsson}, \citenamefont {Bortoli}, \citenamefont {Bourassa}, \citenamefont {Bovaird}, \citenamefont {Brill}, \citenamefont {Broughton}, \citenamefont {Buckley}, \citenamefont {Buell}, \citenamefont {Burger}, \citenamefont {Burkett}, \citenamefont {Bushnell}, \citenamefont {Chen}, \citenamefont {Chiaro}, \citenamefont {Chik}, \citenamefont {Chou}, \citenamefont {Cogan}, \citenamefont {Collins}, \citenamefont {Conner}, \citenamefont {Courtney}, \citenamefont {Crook}, \citenamefont {Curtin}, \citenamefont {Dau}, \citenamefont {Debroy}, \citenamefont {Barba},
  \citenamefont {Demura}, \citenamefont {Di~Paolo}, \citenamefont {Drozdov}, \citenamefont {Dunsworth}, \citenamefont {Erickson}, \citenamefont {Faoro}, \citenamefont {Farhi}, \citenamefont {Fatemi}, \citenamefont {Ferreira}, \citenamefont {Forati}, \citenamefont {Fowler}, \citenamefont {Foxen}, \citenamefont {Genois}, \citenamefont {Giang}, \citenamefont {Gidney}, \citenamefont {Gilboa}, \citenamefont {Giustina}, \citenamefont {Gosula}, \citenamefont {Gross}, \citenamefont {Habegger}, \citenamefont {Hamilton}, \citenamefont {Hansen}, \citenamefont {Harrigan}, \citenamefont {Harrington}, \citenamefont {Heu}, \citenamefont {Hoffmann}, \citenamefont {Hong}, \citenamefont {Huang}, \citenamefont {Huff}, \citenamefont {Huggins}, \citenamefont {Ioffe}, \citenamefont {Isakov}, \citenamefont {Iveland}, \citenamefont {Jeffrey}, \citenamefont {Jiang}, \citenamefont {Jones}, \citenamefont {Juhas}, \citenamefont {Kafri}, \citenamefont {Kechedzhi}, \citenamefont {Khattar}, \citenamefont {Khezri}, \citenamefont
  {Kieferova}, \citenamefont {Kim}, \citenamefont {Kitaev}, \citenamefont {Klots}, \citenamefont {Korotkov}, \citenamefont {Kostritsa}, \citenamefont {Kreikebaum}, \citenamefont {Landhuis}, \citenamefont {Laptev}, \citenamefont {Lau}, \citenamefont {Laws}, \citenamefont {Lee}, \citenamefont {Lee}, \citenamefont {Lensky}, \citenamefont {Lester}, \citenamefont {Lill}, \citenamefont {Liu}, \citenamefont {Locharla}, \citenamefont {Malone}, \citenamefont {Martin}, \citenamefont {McClean}, \citenamefont {McEwen}, \citenamefont {Mieszala}, \citenamefont {Montazeri}, \citenamefont {Morvan}, \citenamefont {Movassagh}, \citenamefont {Mruczkiewicz}, \citenamefont {Neeley}, \citenamefont {Neill}, \citenamefont {Nersisyan}, \citenamefont {Newman}, \citenamefont {Ng}, \citenamefont {Nguyen}, \citenamefont {Nguyen}, \citenamefont {Niu}, \citenamefont {OBrien}, \citenamefont {Opremcak}, \citenamefont {Petukhov}, \citenamefont {Potter}, \citenamefont {Pryadko}, \citenamefont {Quintana}, \citenamefont {Rocque}, \citenamefont
  {Rubin}, \citenamefont {Saei}, \citenamefont {Sank}, \citenamefont {Sankaragomathi}, \citenamefont {Satzinger}, \citenamefont {Schurkus}, \citenamefont {Schuster}, \citenamefont {Shearn}, \citenamefont {Shorter}, \citenamefont {Shutty}, \citenamefont {Shvarts}, \citenamefont {Skruzny}, \citenamefont {Smith}, \citenamefont {Somma}, \citenamefont {Sterling}, \citenamefont {Strain}, \citenamefont {Szalay}, \citenamefont {Torres}, \citenamefont {Vidal}, \citenamefont {Villalonga}, \citenamefont {Heidweiller}, \citenamefont {White}, \citenamefont {Woo}, \citenamefont {Xing}, \citenamefont {Yao}, \citenamefont {Yeh}, \citenamefont {Yoo}, \citenamefont {Young}, \citenamefont {Zalcman}, \citenamefont {Zhang}, \citenamefont {Zhu}, \citenamefont {Zobrist}, \citenamefont {Neven}, \citenamefont {Babbush}, \citenamefont {Bacon}, \citenamefont {Boixo}, \citenamefont {Hilton}, \citenamefont {Lucero}, \citenamefont {Megrant}, \citenamefont {Kelly}, \citenamefont {Chen}, \citenamefont {Roushan}, \citenamefont
  {Smelyanskiy},\ and\ \citenamefont {Abanin}}]{miStableQuantumCorrelatedMany2023}%
  \BibitemOpen
  \bibfield  {author} {\bibinfo {author} {\bibfnamefont {X.}~\bibnamefont {Mi}}, \bibinfo {author} {\bibfnamefont {A.~A.}\ \bibnamefont {Michailidis}}, \bibinfo {author} {\bibfnamefont {S.}~\bibnamefont {Shabani}}, \bibinfo {author} {\bibfnamefont {K.~C.}\ \bibnamefont {Miao}}, \bibinfo {author} {\bibfnamefont {P.~V.}\ \bibnamefont {Klimov}}, \bibinfo {author} {\bibfnamefont {J.}~\bibnamefont {Lloyd}}, \bibinfo {author} {\bibfnamefont {E.}~\bibnamefont {Rosenberg}}, \bibinfo {author} {\bibfnamefont {R.}~\bibnamefont {Acharya}}, \bibinfo {author} {\bibfnamefont {I.}~\bibnamefont {Aleiner}}, \bibinfo {author} {\bibfnamefont {T.~I.}\ \bibnamefont {Andersen}}, \bibinfo {author} {\bibfnamefont {M.}~\bibnamefont {Ansmann}}, \bibinfo {author} {\bibfnamefont {F.}~\bibnamefont {Arute}}, \bibinfo {author} {\bibfnamefont {K.}~\bibnamefont {Arya}}, \bibinfo {author} {\bibfnamefont {A.}~\bibnamefont {Asfaw}}, \bibinfo {author} {\bibfnamefont {J.}~\bibnamefont {Atalaya}}, \bibinfo {author} {\bibfnamefont {J.~C.}\
  \bibnamefont {Bardin}}, \bibinfo {author} {\bibfnamefont {A.}~\bibnamefont {Bengtsson}}, \bibinfo {author} {\bibfnamefont {G.}~\bibnamefont {Bortoli}}, \bibinfo {author} {\bibfnamefont {A.}~\bibnamefont {Bourassa}}, \bibinfo {author} {\bibfnamefont {J.}~\bibnamefont {Bovaird}}, \bibinfo {author} {\bibfnamefont {L.}~\bibnamefont {Brill}}, \bibinfo {author} {\bibfnamefont {M.}~\bibnamefont {Broughton}}, \bibinfo {author} {\bibfnamefont {B.~B.}\ \bibnamefont {Buckley}}, \bibinfo {author} {\bibfnamefont {D.~A.}\ \bibnamefont {Buell}}, \bibinfo {author} {\bibfnamefont {T.}~\bibnamefont {Burger}}, \bibinfo {author} {\bibfnamefont {B.}~\bibnamefont {Burkett}}, \bibinfo {author} {\bibfnamefont {N.}~\bibnamefont {Bushnell}}, \bibinfo {author} {\bibfnamefont {Z.}~\bibnamefont {Chen}}, \bibinfo {author} {\bibfnamefont {B.}~\bibnamefont {Chiaro}}, \bibinfo {author} {\bibfnamefont {D.}~\bibnamefont {Chik}}, \bibinfo {author} {\bibfnamefont {C.}~\bibnamefont {Chou}}, \bibinfo {author} {\bibfnamefont {J.}~\bibnamefont
  {Cogan}}, \bibinfo {author} {\bibfnamefont {R.}~\bibnamefont {Collins}}, \bibinfo {author} {\bibfnamefont {P.}~\bibnamefont {Conner}}, \bibinfo {author} {\bibfnamefont {W.}~\bibnamefont {Courtney}}, \bibinfo {author} {\bibfnamefont {A.~L.}\ \bibnamefont {Crook}}, \bibinfo {author} {\bibfnamefont {B.}~\bibnamefont {Curtin}}, \bibinfo {author} {\bibfnamefont {A.~G.}\ \bibnamefont {Dau}}, \bibinfo {author} {\bibfnamefont {D.~M.}\ \bibnamefont {Debroy}}, \bibinfo {author} {\bibfnamefont {A.~D.~T.}\ \bibnamefont {Barba}}, \bibinfo {author} {\bibfnamefont {S.}~\bibnamefont {Demura}}, \bibinfo {author} {\bibfnamefont {A.}~\bibnamefont {Di~Paolo}}, \bibinfo {author} {\bibfnamefont {I.~K.}\ \bibnamefont {Drozdov}}, \bibinfo {author} {\bibfnamefont {A.}~\bibnamefont {Dunsworth}}, \bibinfo {author} {\bibfnamefont {C.}~\bibnamefont {Erickson}}, \bibinfo {author} {\bibfnamefont {L.}~\bibnamefont {Faoro}}, \bibinfo {author} {\bibfnamefont {E.}~\bibnamefont {Farhi}}, \bibinfo {author} {\bibfnamefont {R.}~\bibnamefont
  {Fatemi}}, \bibinfo {author} {\bibfnamefont {V.~S.}\ \bibnamefont {Ferreira}}, \bibinfo {author} {\bibfnamefont {L.~F. B.~E.}\ \bibnamefont {Forati}}, \bibinfo {author} {\bibfnamefont {A.~G.}\ \bibnamefont {Fowler}}, \bibinfo {author} {\bibfnamefont {B.}~\bibnamefont {Foxen}}, \bibinfo {author} {\bibfnamefont {E.}~\bibnamefont {Genois}}, \bibinfo {author} {\bibfnamefont {W.}~\bibnamefont {Giang}}, \bibinfo {author} {\bibfnamefont {C.}~\bibnamefont {Gidney}}, \bibinfo {author} {\bibfnamefont {D.}~\bibnamefont {Gilboa}}, \bibinfo {author} {\bibfnamefont {M.}~\bibnamefont {Giustina}}, \bibinfo {author} {\bibfnamefont {R.}~\bibnamefont {Gosula}}, \bibinfo {author} {\bibfnamefont {J.~A.}\ \bibnamefont {Gross}}, \bibinfo {author} {\bibfnamefont {S.}~\bibnamefont {Habegger}}, \bibinfo {author} {\bibfnamefont {M.~C.}\ \bibnamefont {Hamilton}}, \bibinfo {author} {\bibfnamefont {M.}~\bibnamefont {Hansen}}, \bibinfo {author} {\bibfnamefont {M.~P.}\ \bibnamefont {Harrigan}}, \bibinfo {author} {\bibfnamefont {S.~D.}\
  \bibnamefont {Harrington}}, \bibinfo {author} {\bibfnamefont {P.}~\bibnamefont {Heu}}, \bibinfo {author} {\bibfnamefont {M.~R.}\ \bibnamefont {Hoffmann}}, \bibinfo {author} {\bibfnamefont {S.}~\bibnamefont {Hong}}, \bibinfo {author} {\bibfnamefont {T.}~\bibnamefont {Huang}}, \bibinfo {author} {\bibfnamefont {A.}~\bibnamefont {Huff}}, \bibinfo {author} {\bibfnamefont {W.~J.}\ \bibnamefont {Huggins}}, \bibinfo {author} {\bibfnamefont {L.~B.}\ \bibnamefont {Ioffe}}, \bibinfo {author} {\bibfnamefont {S.~V.}\ \bibnamefont {Isakov}}, \bibinfo {author} {\bibfnamefont {J.}~\bibnamefont {Iveland}}, \bibinfo {author} {\bibfnamefont {E.}~\bibnamefont {Jeffrey}}, \bibinfo {author} {\bibfnamefont {Z.}~\bibnamefont {Jiang}}, \bibinfo {author} {\bibfnamefont {C.}~\bibnamefont {Jones}}, \bibinfo {author} {\bibfnamefont {P.}~\bibnamefont {Juhas}}, \bibinfo {author} {\bibfnamefont {D.}~\bibnamefont {Kafri}}, \bibinfo {author} {\bibfnamefont {K.}~\bibnamefont {Kechedzhi}}, \bibinfo {author} {\bibfnamefont {T.}~\bibnamefont
  {Khattar}}, \bibinfo {author} {\bibfnamefont {M.}~\bibnamefont {Khezri}}, \bibinfo {author} {\bibfnamefont {M.}~\bibnamefont {Kieferova}}, \bibinfo {author} {\bibfnamefont {S.}~\bibnamefont {Kim}}, \bibinfo {author} {\bibfnamefont {A.}~\bibnamefont {Kitaev}}, \bibinfo {author} {\bibfnamefont {A.~R.}\ \bibnamefont {Klots}}, \bibinfo {author} {\bibfnamefont {A.~N.}\ \bibnamefont {Korotkov}}, \bibinfo {author} {\bibfnamefont {F.}~\bibnamefont {Kostritsa}}, \bibinfo {author} {\bibfnamefont {J.~M.}\ \bibnamefont {Kreikebaum}}, \bibinfo {author} {\bibfnamefont {D.}~\bibnamefont {Landhuis}}, \bibinfo {author} {\bibfnamefont {P.}~\bibnamefont {Laptev}}, \bibinfo {author} {\bibfnamefont {K.-M.}\ \bibnamefont {Lau}}, \bibinfo {author} {\bibfnamefont {L.}~\bibnamefont {Laws}}, \bibinfo {author} {\bibfnamefont {J.}~\bibnamefont {Lee}}, \bibinfo {author} {\bibfnamefont {K.~W.}\ \bibnamefont {Lee}}, \bibinfo {author} {\bibfnamefont {Y.~D.}\ \bibnamefont {Lensky}}, \bibinfo {author} {\bibfnamefont {B.~J.}\ \bibnamefont
  {Lester}}, \bibinfo {author} {\bibfnamefont {A.~T.}\ \bibnamefont {Lill}}, \bibinfo {author} {\bibfnamefont {W.}~\bibnamefont {Liu}}, \bibinfo {author} {\bibfnamefont {A.}~\bibnamefont {Locharla}}, \bibinfo {author} {\bibfnamefont {F.~D.}\ \bibnamefont {Malone}}, \bibinfo {author} {\bibfnamefont {O.}~\bibnamefont {Martin}}, \bibinfo {author} {\bibfnamefont {J.~R.}\ \bibnamefont {McClean}}, \bibinfo {author} {\bibfnamefont {M.}~\bibnamefont {McEwen}}, \bibinfo {author} {\bibfnamefont {A.}~\bibnamefont {Mieszala}}, \bibinfo {author} {\bibfnamefont {S.}~\bibnamefont {Montazeri}}, \bibinfo {author} {\bibfnamefont {A.}~\bibnamefont {Morvan}}, \bibinfo {author} {\bibfnamefont {R.}~\bibnamefont {Movassagh}}, \bibinfo {author} {\bibfnamefont {W.}~\bibnamefont {Mruczkiewicz}}, \bibinfo {author} {\bibfnamefont {M.}~\bibnamefont {Neeley}}, \bibinfo {author} {\bibfnamefont {C.}~\bibnamefont {Neill}}, \bibinfo {author} {\bibfnamefont {A.}~\bibnamefont {Nersisyan}}, \bibinfo {author} {\bibfnamefont {M.}~\bibnamefont
  {Newman}}, \bibinfo {author} {\bibfnamefont {J.~H.}\ \bibnamefont {Ng}}, \bibinfo {author} {\bibfnamefont {A.}~\bibnamefont {Nguyen}}, \bibinfo {author} {\bibfnamefont {M.}~\bibnamefont {Nguyen}}, \bibinfo {author} {\bibfnamefont {M.~Y.}\ \bibnamefont {Niu}}, \bibinfo {author} {\bibfnamefont {T.~E.}\ \bibnamefont {OBrien}}, \bibinfo {author} {\bibfnamefont {A.}~\bibnamefont {Opremcak}}, \bibinfo {author} {\bibfnamefont {A.}~\bibnamefont {Petukhov}}, \bibinfo {author} {\bibfnamefont {R.}~\bibnamefont {Potter}}, \bibinfo {author} {\bibfnamefont {L.~P.}\ \bibnamefont {Pryadko}}, \bibinfo {author} {\bibfnamefont {C.}~\bibnamefont {Quintana}}, \bibinfo {author} {\bibfnamefont {C.}~\bibnamefont {Rocque}}, \bibinfo {author} {\bibfnamefont {N.~C.}\ \bibnamefont {Rubin}}, \bibinfo {author} {\bibfnamefont {N.}~\bibnamefont {Saei}}, \bibinfo {author} {\bibfnamefont {D.}~\bibnamefont {Sank}}, \bibinfo {author} {\bibfnamefont {K.}~\bibnamefont {Sankaragomathi}}, \bibinfo {author} {\bibfnamefont {K.~J.}\ \bibnamefont
  {Satzinger}}, \bibinfo {author} {\bibfnamefont {H.~F.}\ \bibnamefont {Schurkus}}, \bibinfo {author} {\bibfnamefont {C.}~\bibnamefont {Schuster}}, \bibinfo {author} {\bibfnamefont {M.~J.}\ \bibnamefont {Shearn}}, \bibinfo {author} {\bibfnamefont {A.}~\bibnamefont {Shorter}}, \bibinfo {author} {\bibfnamefont {N.}~\bibnamefont {Shutty}}, \bibinfo {author} {\bibfnamefont {V.}~\bibnamefont {Shvarts}}, \bibinfo {author} {\bibfnamefont {J.}~\bibnamefont {Skruzny}}, \bibinfo {author} {\bibfnamefont {W.~C.}\ \bibnamefont {Smith}}, \bibinfo {author} {\bibfnamefont {R.}~\bibnamefont {Somma}}, \bibinfo {author} {\bibfnamefont {G.}~\bibnamefont {Sterling}}, \bibinfo {author} {\bibfnamefont {D.}~\bibnamefont {Strain}}, \bibinfo {author} {\bibfnamefont {M.}~\bibnamefont {Szalay}}, \bibinfo {author} {\bibfnamefont {A.}~\bibnamefont {Torres}}, \bibinfo {author} {\bibfnamefont {G.}~\bibnamefont {Vidal}}, \bibinfo {author} {\bibfnamefont {B.}~\bibnamefont {Villalonga}}, \bibinfo {author} {\bibfnamefont {C.~V.}\ \bibnamefont
  {Heidweiller}}, \bibinfo {author} {\bibfnamefont {T.}~\bibnamefont {White}}, \bibinfo {author} {\bibfnamefont {B.~W.~K.}\ \bibnamefont {Woo}}, \bibinfo {author} {\bibfnamefont {C.}~\bibnamefont {Xing}}, \bibinfo {author} {\bibfnamefont {Z.~J.}\ \bibnamefont {Yao}}, \bibinfo {author} {\bibfnamefont {P.}~\bibnamefont {Yeh}}, \bibinfo {author} {\bibfnamefont {J.}~\bibnamefont {Yoo}}, \bibinfo {author} {\bibfnamefont {G.}~\bibnamefont {Young}}, \bibinfo {author} {\bibfnamefont {A.}~\bibnamefont {Zalcman}}, \bibinfo {author} {\bibfnamefont {Y.}~\bibnamefont {Zhang}}, \bibinfo {author} {\bibfnamefont {N.}~\bibnamefont {Zhu}}, \bibinfo {author} {\bibfnamefont {N.}~\bibnamefont {Zobrist}}, \bibinfo {author} {\bibfnamefont {H.}~\bibnamefont {Neven}}, \bibinfo {author} {\bibfnamefont {R.}~\bibnamefont {Babbush}}, \bibinfo {author} {\bibfnamefont {D.}~\bibnamefont {Bacon}}, \bibinfo {author} {\bibfnamefont {S.}~\bibnamefont {Boixo}}, \bibinfo {author} {\bibfnamefont {J.}~\bibnamefont {Hilton}}, \bibinfo {author}
  {\bibfnamefont {E.}~\bibnamefont {Lucero}}, \bibinfo {author} {\bibfnamefont {A.}~\bibnamefont {Megrant}}, \bibinfo {author} {\bibfnamefont {J.}~\bibnamefont {Kelly}}, \bibinfo {author} {\bibfnamefont {Y.}~\bibnamefont {Chen}}, \bibinfo {author} {\bibfnamefont {P.}~\bibnamefont {Roushan}}, \bibinfo {author} {\bibfnamefont {V.}~\bibnamefont {Smelyanskiy}},\ and\ \bibinfo {author} {\bibfnamefont {D.~A.}\ \bibnamefont {Abanin}},\ }\href {https://doi.org/10.48550/arXiv.2304.13878} {\bibinfo {title} {Stable {{Quantum-Correlated Many Body States}} via {{Engineered Dissipation}}}} (\bibinfo {year} {2023}),\ \Eprint {https://arxiv.org/abs/2304.13878} {arxiv:2304.13878 [quant-ph]} \BibitemShut {NoStop}%
\bibitem [{\citenamefont {Wang}\ \emph {et~al.}(2023)\citenamefont {Wang}, \citenamefont {Snizhko}, \citenamefont {Romito}, \citenamefont {Gefen},\ and\ \citenamefont {Murch}}]{wangDissipativePreparationStabilization2023}%
  \BibitemOpen
  \bibfield  {author} {\bibinfo {author} {\bibfnamefont {Y.}~\bibnamefont {Wang}}, \bibinfo {author} {\bibfnamefont {K.}~\bibnamefont {Snizhko}}, \bibinfo {author} {\bibfnamefont {A.}~\bibnamefont {Romito}}, \bibinfo {author} {\bibfnamefont {Y.}~\bibnamefont {Gefen}},\ and\ \bibinfo {author} {\bibfnamefont {K.}~\bibnamefont {Murch}},\ }\bibfield  {title} {\bibinfo {title} {Dissipative preparation and stabilization of many-body quantum states in a superconducting qutrit array},\ }\href {https://doi.org/10.1103/PhysRevA.108.013712} {\bibfield  {journal} {\bibinfo  {journal} {Physical Review A}\ }\textbf {\bibinfo {volume} {108}},\ \bibinfo {pages} {013712} (\bibinfo {year} {2023})},\ \Eprint {https://arxiv.org/abs/2303.12111} {arxiv:2303.12111 [cond-mat, physics:quant-ph]} \BibitemShut {NoStop}%
\bibitem [{\citenamefont {Metelmann}\ and\ \citenamefont {T{\"u}reci}(2018)}]{metelmannNonreciprocalSignalRouting2018}%
  \BibitemOpen
  \bibfield  {author} {\bibinfo {author} {\bibfnamefont {A.}~\bibnamefont {Metelmann}}\ and\ \bibinfo {author} {\bibfnamefont {H.~E.}\ \bibnamefont {T{\"u}reci}},\ }\bibfield  {title} {\bibinfo {title} {Nonreciprocal signal routing in an active quantum network},\ }\href {https://doi.org/10.1103/PhysRevA.97.043833} {\bibfield  {journal} {\bibinfo  {journal} {Physical Review A}\ }\textbf {\bibinfo {volume} {97}},\ \bibinfo {pages} {043833} (\bibinfo {year} {2018})}\BibitemShut {NoStop}%
\bibitem [{\citenamefont {Ma}\ \emph {et~al.}(2019)\citenamefont {Ma}, \citenamefont {Li}, \citenamefont {Liu}, \citenamefont {Xie},\ and\ \citenamefont {Li}}]{maStabilizingBellStates2019}%
  \BibitemOpen
  \bibfield  {author} {\bibinfo {author} {\bibfnamefont {S.-l.}\ \bibnamefont {Ma}}, \bibinfo {author} {\bibfnamefont {X.-k.}\ \bibnamefont {Li}}, \bibinfo {author} {\bibfnamefont {X.-y.}\ \bibnamefont {Liu}}, \bibinfo {author} {\bibfnamefont {J.-k.}\ \bibnamefont {Xie}},\ and\ \bibinfo {author} {\bibfnamefont {F.-l.}\ \bibnamefont {Li}},\ }\bibfield  {title} {\bibinfo {title} {Stabilizing {{Bell}} states of two separated superconducting qubits via quantum reservoir engineering},\ }\href {https://doi.org/10.1103/PhysRevA.99.042336} {\bibfield  {journal} {\bibinfo  {journal} {Physical Review A}\ }\textbf {\bibinfo {volume} {99}},\ \bibinfo {pages} {042336} (\bibinfo {year} {2019})}\BibitemShut {NoStop}%
\bibitem [{\citenamefont {Potts}\ \emph {et~al.}(2001)\citenamefont {Potts}, \citenamefont {Parker}, \citenamefont {Baumberg},\ and\ \citenamefont {de~Groot}}]{pottsCMOSCompatibleFabrication2001}%
  \BibitemOpen
  \bibfield  {author} {\bibinfo {author} {\bibfnamefont {A.}~\bibnamefont {Potts}}, \bibinfo {author} {\bibfnamefont {G.~J.}\ \bibnamefont {Parker}}, \bibinfo {author} {\bibfnamefont {J.~J.}\ \bibnamefont {Baumberg}},\ and\ \bibinfo {author} {\bibfnamefont {P.~A.~J.}\ \bibnamefont {de~Groot}},\ }\bibfield  {title} {\bibinfo {title} {{{CMOS}} compatible fabrication methods for submicron {{Josephson}} junction qubits},\ }\href {https://doi.org/10.1049/ip-smt:20010395} {\bibfield  {journal} {\bibinfo  {journal} {IEE Proceedings - Science, Measurement and Technology}\ }\textbf {\bibinfo {volume} {148}},\ \bibinfo {pages} {225} (\bibinfo {year} {2001})}\BibitemShut {NoStop}%
\bibitem [{\citenamefont {Costache}\ \emph {et~al.}(2012)\citenamefont {Costache}, \citenamefont {Bridoux}, \citenamefont {Neumann},\ and\ \citenamefont {Valenzuela}}]{costacheLateralMetallicDevices2012}%
  \BibitemOpen
  \bibfield  {author} {\bibinfo {author} {\bibfnamefont {M.~V.}\ \bibnamefont {Costache}}, \bibinfo {author} {\bibfnamefont {G.}~\bibnamefont {Bridoux}}, \bibinfo {author} {\bibfnamefont {I.}~\bibnamefont {Neumann}},\ and\ \bibinfo {author} {\bibfnamefont {S.~O.}\ \bibnamefont {Valenzuela}},\ }\bibfield  {title} {\bibinfo {title} {Lateral metallic devices made by a multiangle shadow evaporation technique},\ }\href {https://doi.org/10.1116/1.4722982} {\bibfield  {journal} {\bibinfo  {journal} {Journal of Vacuum Science \& Technology B}\ }\textbf {\bibinfo {volume} {30}},\ \bibinfo {pages} {04E105} (\bibinfo {year} {2012})}\BibitemShut {NoStop}%
\bibitem [{\citenamefont {Koch}\ \emph {et~al.}(2007)\citenamefont {Koch}, \citenamefont {Yu}, \citenamefont {Gambetta}, \citenamefont {Houck}, \citenamefont {Schuster}, \citenamefont {Majer}, \citenamefont {Blais}, \citenamefont {Devoret}, \citenamefont {Girvin},\ and\ \citenamefont {Schoelkopf}}]{kochChargeinsensitiveQubitDesign2007}%
  \BibitemOpen
  \bibfield  {author} {\bibinfo {author} {\bibfnamefont {J.}~\bibnamefont {Koch}}, \bibinfo {author} {\bibfnamefont {T.}~\bibnamefont {Yu}}, \bibinfo {author} {\bibfnamefont {J.}~\bibnamefont {Gambetta}}, \bibinfo {author} {\bibfnamefont {a.}~\bibnamefont {Houck}}, \bibinfo {author} {\bibfnamefont {D.}~\bibnamefont {Schuster}}, \bibinfo {author} {\bibfnamefont {J.}~\bibnamefont {Majer}}, \bibinfo {author} {\bibfnamefont {A.}~\bibnamefont {Blais}}, \bibinfo {author} {\bibfnamefont {M.}~\bibnamefont {Devoret}}, \bibinfo {author} {\bibfnamefont {S.}~\bibnamefont {Girvin}},\ and\ \bibinfo {author} {\bibfnamefont {R.}~\bibnamefont {Schoelkopf}},\ }\bibfield  {title} {\bibinfo {title} {Charge-insensitive qubit design derived from the {{Cooper}} pair box},\ }\href {https://doi.org/10.1103/PhysRevA.76.042319} {\bibfield  {journal} {\bibinfo  {journal} {Physical Review A}\ }\textbf {\bibinfo {volume} {76}},\ \bibinfo {pages} {1} (\bibinfo {year} {2007})}\BibitemShut {NoStop}%
\bibitem [{\citenamefont {Jin}\ \emph {et~al.}(2015)\citenamefont {Jin}, \citenamefont {Kamal}, \citenamefont {Sears}, \citenamefont {Gudmundsen}, \citenamefont {Hover}, \citenamefont {Miloshi}, \citenamefont {Slattery}, \citenamefont {Yan}, \citenamefont {Yoder}, \citenamefont {Orlando}, \citenamefont {Gustavsson},\ and\ \citenamefont {Oliver}}]{jinThermalResidualExcitedState2015}%
  \BibitemOpen
  \bibfield  {author} {\bibinfo {author} {\bibfnamefont {X.~Y.}\ \bibnamefont {Jin}}, \bibinfo {author} {\bibfnamefont {A.}~\bibnamefont {Kamal}}, \bibinfo {author} {\bibfnamefont {A.~P.}\ \bibnamefont {Sears}}, \bibinfo {author} {\bibfnamefont {T.}~\bibnamefont {Gudmundsen}}, \bibinfo {author} {\bibfnamefont {D.}~\bibnamefont {Hover}}, \bibinfo {author} {\bibfnamefont {J.}~\bibnamefont {Miloshi}}, \bibinfo {author} {\bibfnamefont {R.}~\bibnamefont {Slattery}}, \bibinfo {author} {\bibfnamefont {F.}~\bibnamefont {Yan}}, \bibinfo {author} {\bibfnamefont {J.}~\bibnamefont {Yoder}}, \bibinfo {author} {\bibfnamefont {T.~P.}\ \bibnamefont {Orlando}}, \bibinfo {author} {\bibfnamefont {S.}~\bibnamefont {Gustavsson}},\ and\ \bibinfo {author} {\bibfnamefont {W.~D.}\ \bibnamefont {Oliver}},\ }\bibfield  {title} {\bibinfo {title} {Thermal and {{Residual Excited-State Population}} in a {{3D Transmon Qubit}}},\ }\href {https://doi.org/10.1103/PhysRevLett.114.240501} {\bibfield  {journal} {\bibinfo  {journal} {Physical
  Review Letters}\ }\textbf {\bibinfo {volume} {114}},\ \bibinfo {pages} {240501} (\bibinfo {year} {2015})}\BibitemShut {NoStop}%
\bibitem [{\citenamefont {Serniak}\ \emph {et~al.}(2018)\citenamefont {Serniak}, \citenamefont {Hays}, \citenamefont {{de Lange}}, \citenamefont {Diamond}, \citenamefont {Shankar}, \citenamefont {Burkhart}, \citenamefont {Frunzio}, \citenamefont {Houzet},\ and\ \citenamefont {Devoret}}]{serniakHotNonequilibriumQuasiparticles2018a}%
  \BibitemOpen
  \bibfield  {author} {\bibinfo {author} {\bibfnamefont {K.}~\bibnamefont {Serniak}}, \bibinfo {author} {\bibfnamefont {M.}~\bibnamefont {Hays}}, \bibinfo {author} {\bibfnamefont {G.}~\bibnamefont {{de Lange}}}, \bibinfo {author} {\bibfnamefont {S.}~\bibnamefont {Diamond}}, \bibinfo {author} {\bibfnamefont {S.}~\bibnamefont {Shankar}}, \bibinfo {author} {\bibfnamefont {L.~D.}\ \bibnamefont {Burkhart}}, \bibinfo {author} {\bibfnamefont {L.}~\bibnamefont {Frunzio}}, \bibinfo {author} {\bibfnamefont {M.}~\bibnamefont {Houzet}},\ and\ \bibinfo {author} {\bibfnamefont {M.~H.}\ \bibnamefont {Devoret}},\ }\bibfield  {title} {\bibinfo {title} {Hot {{Nonequilibrium Quasiparticles}} in {{Transmon Qubits}}},\ }\href {https://doi.org/10.1103/PhysRevLett.121.157701} {\bibfield  {journal} {\bibinfo  {journal} {Physical Review Letters}\ }\textbf {\bibinfo {volume} {121}},\ \bibinfo {pages} {157701} (\bibinfo {year} {2018})}\BibitemShut {NoStop}%
\end{thebibliography}%
\end{document}